\documentclass[aip, amsmath, amssymb, reprint]{revtex4-1}

\usepackage{graphicx} 
\usepackage{dcolumn}  
\usepackage{bm}       

\usepackage[utf8]{inputenc}
\usepackage[T1]{fontenc}
\usepackage{mathptmx}
\usepackage{etoolbox}

\makeatletter
\def\@email#1#2{%
 \endgroup
 \patchcmd{\titleblock@produce}
  {\frontmatter@RRAPformat}
  {\frontmatter@RRAPformat{\produce@RRAP{*#1\href{mailto:#2}{#2}}}\frontmatter@RRAPformat}
  {}{}
}%
\makeatother
\begin{document}

\preprint{AIP/123-QED}

\title{Enabling Structure-Only Initialization and Out-of-Distribution Generalization in GNN-based Molecular Dynamics Simulators} 

\author{S. A. Shteingolts}
\author{Salman N. Salman}%
\author{Dan Mendels}
\email{danmendels@technion.ac.il}
\affiliation{The Wolfson Department of Chemical Engineering, Technion - Israel Institute of Technology, Haifa, 32000, Israel}

\date{\today}

\begin{abstract}
Machine learning-based simulators offer the potential to model the dynamics of complex systems significantly more efficiently than classical approaches, while retaining differentiability, a key property for materials design and optimization. In recent years, graph neural network (GNN)-based simulators have shown strong performance across a range of physical domains, including molecular dynamics. However, their reliance on temporal context, i.e., sequences of previous states, for accurate prediction can limit their applicability in inverse design settings, where simulations must be initialized from a single structural configuration. Moreover, inverse design inherently requires robust out-of-distribution (OOD) generalization, as candidate structures typically lie outside the training domain. Here, we address both challenges by introducing two complementary strategies that enable stable and accurate structure-only initialization of GNN-based simulations for previously unseen systems. To explicitly target OOD generalization, we further propose an inference-time physics-based optimization framework that constrains model predictions to remain physically consistent during rollout. In addition, we introduce a differentiable, GNN-based barostat that enables accurate tracking of system dimensions and pressure, which is critical for capturing macroscopic responses and supporting generalization in extrapolative regimes. We evaluate these approaches in the context of uniaxial compression of two-dimensional disordered elastic networks spanning a broad range of geometries, Poisson ratios, and microscopic behaviors. We find that, in combination, these methods substantially improve rollout stability and enable reliable OOD generalization, including to regimes exhibiting qualitatively distinct and more complex dynamics than those represented in the training data. These results demonstrate that, when appropriately initialized and constrained, GNN-based simulators can serve as efficient and generalizable tools for materials discovery and structural optimization, marking a step toward their broader use in materials, molecular, and dynamical system design and engineering.

\end{abstract}

\maketitle

\section{Introduction}\label{sec::intro}

\begin{figure*}[t]
    \centering
    \includegraphics{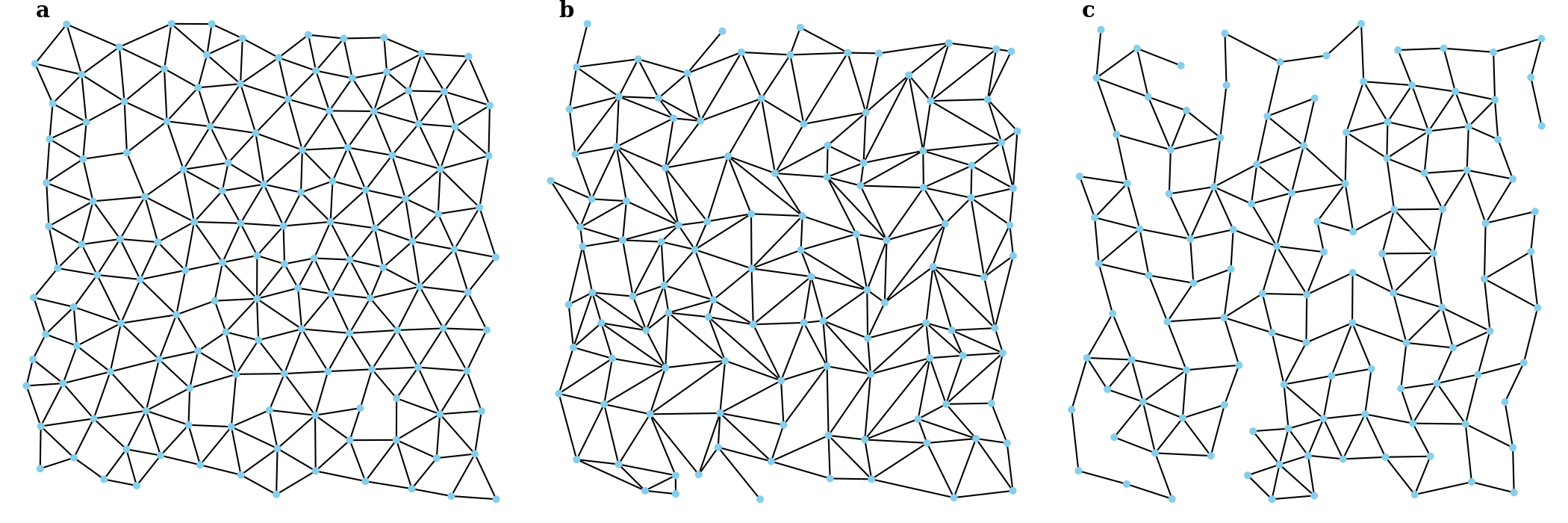}
    \caption{Examples of disordered elastic network (DEN) topologies. (a) - a non-auxetic DEN with high Poisson ratio, (b) - auxetic DEN optimized via global node optimization, and (c) - auxetic DEN from direct bond stiffness optimization.}
    \label{fig::dens}
\end{figure*}

When simulating dynamical systems, machine learning (ML) offers a computationally efficient alternative to traditional numerical solvers while retaining end-to-end differentiability.\cite{Sanchez-Gonzalez2020, Pfaff2021, Salman2025EvaluatingNetworks, Iparraguirre2026MeshGraphNet-Transformer:Mechanics} This differentiability provides a powerful framework for inverse design, enabling direct optimization of system dynamics toward desired behaviors and functionalities.\cite{Zheng2021, Wang2022, Allen2022, Tian2022, Zheng2023}

Despite recent progress, a central challenge in applying ML-based simulators to inverse design arises from their reliance on temporal context. Accurate prediction of complex dynamics typically requires access to dynamic state variables, such as velocity or acceleration, often across multiple preceding timesteps. While such temporal information improves prediction accuracy and rollout stability, it is generally unavailable in inverse design settings, where simulations must be initialized from a single, static configuration.\cite{Allen2022b, Yang2022LearningNetworks, Salman2025EvaluatingNetworks} Consequently, models that depend on extended temporal histories are fundamentally incompatible with this “cold-start” setting, presenting a substantial barrier for integrating ML-based simulators into gradient-based optimization pipelines.

A second, closely related challenge is out-of-distribution (OOD) generalization. Inverse design inherently requires evaluating candidate systems that lie outside the training distribution. However, implicit structure-property models often rely on large, expensive datasets and exhibit limited generalization beyond the sampled design space.\cite{Mendels2018, Mendels2022a, Mendels2022, Nandy2022AudacityDiscovery, Dou2023MachineScience, Smit2026TheDiscovery, Zhilkin2026GuidingBarriers, Medaparambath2026CollectivePeptide} In contrast, ML-based simulators function as explicit forward models, offering the potential for improved data efficiency and generalization by learning aspects of the underlying physical dynamics rather than static structure-property relationships.\cite{Salman2025EvaluatingNetworks}

In this work, we address both the structure-only initialization (SOI) and OOD generalization challenges. We introduce two complementary strategies for enabling stable and accurate SOI of GNN-based simulators. The first approach employs a custom, minimal, differentiable molecular dynamics (MD) engine to bootstrap the GNN simulator by generating a short initial trajectory that provides the required dynamical context. The second approach is fully data-driven and consists of a cascade of specialized GNN models, each trained to predict a specific timestep, thereby progressively constructing the temporal context required by the main GNN simulator. We show that both approaches enable stable rollouts from a single static configuration, overcoming the limitations of standard GNN-based simulators.

To further improve robustness in extrapolative regimes, we introduce an inference-time physics-based optimization (ITPO) framework, which constrains model predictions to remain physically consistent during rollout. This approach reduces error accumulation and improves generalization, particularly in regimes that deviate significantly from the training distribution.

We evaluate the proposed methods in the context of uniaxial compression of two-dimensional disordered elastic networks (DENs), which exhibit a broad range of geometries, Poisson's ratios, and microscopic behaviors.\cite{Reid2018, Rens2019, Mendels2022, Shen2024} Specifically, we focus on the "cold-start" problem of initiating dynamics from a single static configuration, as well as the model's ability to generalize to elastic networks with substantially lower Poisson’s ratios. These low-$\nu$ regimes serve as a rigorous benchmark for the simulator's extrapolative capabilities, exhibiting distinct and more complex dynamics than those present in the high-$\nu$ training data.

\section{Methods}

\begin{figure*}[t]
    \centering
    \includegraphics{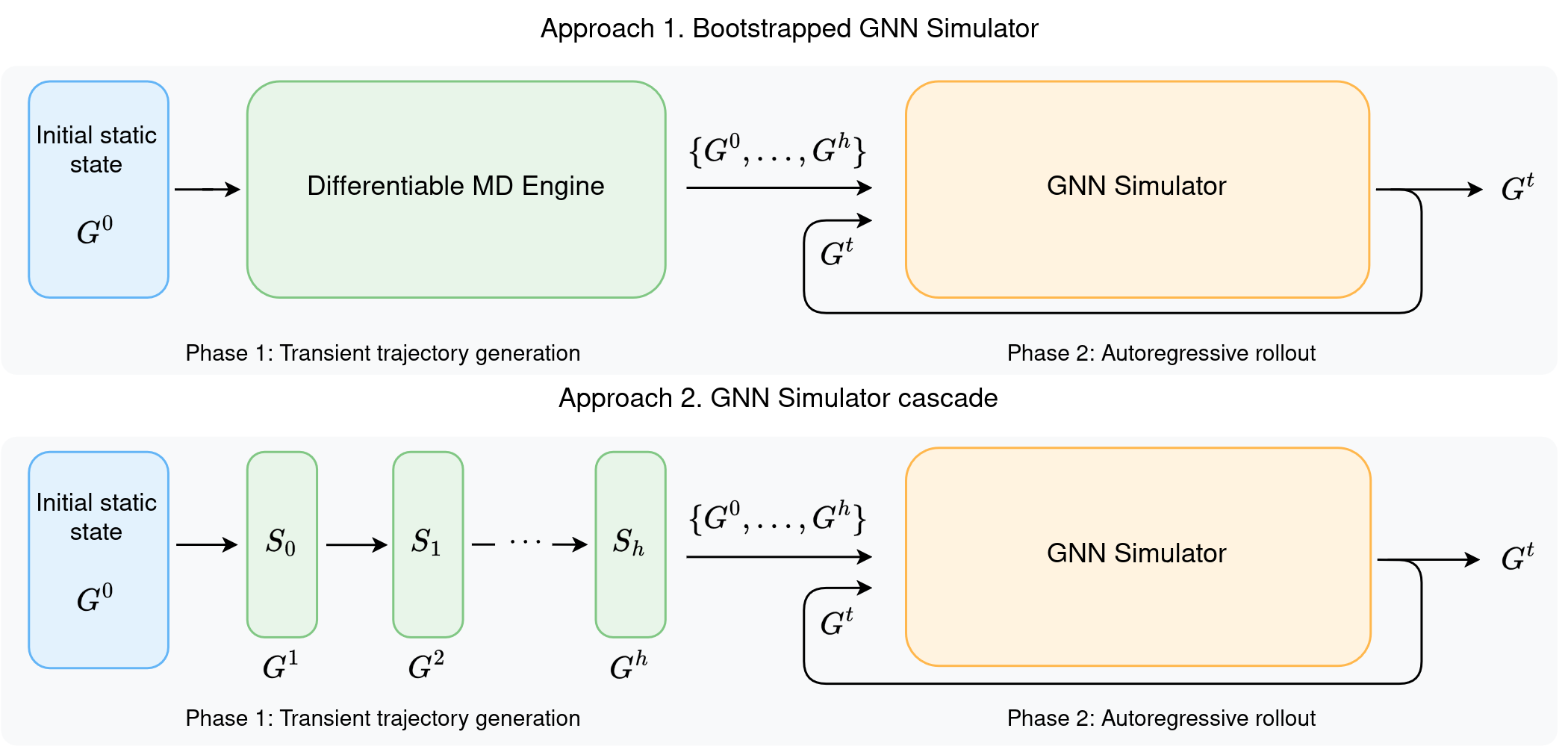}
    \caption{Schematic overview of the two proposed approaches for addressing the SOI problem of GNN-based simulators. The first approach (top) employs a custom differentiable MD engine to generate initial dynamics, while the second approach (bottom) uses a cascade of specialized GNN simulators tailored to the initial timesteps.}
    \label{fig::diagram}
\end{figure*}

\subsection{GNN Simulator}

Our primary tool within this study is a GNN simulator based on an established architectural framework for modeling dynamic systems.\cite{Battaglia2018RelationalNetworks, Sanchez-Gonzalez2020, Pfaff2021, Salman2025EvaluatingNetworks} The simulator is designed to predict the subsequent state in a compression trajectory based on a sequence of preceding configurations. Specifically, the system configuration at step $t$ is encoded into an input graph $G^t$, consisting of nodes $V$ (elastic network beads) and edges $E$ (harmonic bonds), along with their associated node features $\textbf{x}$ and edge features $\textbf{e}$. See Section \ref{sec::comp} for details.

The simulator takes $G^t$ as input and outputs predicted particle accelerations $\hat{\textbf{a}}^t$, which are then used to calculate next-step positions $\textbf{r}^{t+1}$ via semi-implicit forward Euler integration:
\begin{eqnarray}
    \textbf{r}^{t+1} = \textbf{r}^t + \textbf{v}^{t} \Delta t +\hat{\textbf{a}}^t \Delta t ^ 2
\end{eqnarray}
with $\Delta t = 1$. When applied auto-regressively, i.e., the output configuration $\textbf{r}^{t+1}$ is used to construct the input graph $G^{t+1}$ for the subsequent prediction, the procedure generates a "rollout" trajectory. The number of prior system states included in the input graph defines the temporal history $h$. Unless otherwise specified, the models discussed herein utilize history $h=3$, meaning that node features $\textbf{x}$ encode the velocities of the current and two preceding timesteps ($\textbf{v}^t$, $\textbf{v}^{t-1}$, $\textbf{v}^{t-2}$). See Section \ref{sec::methods::mst} for details on the training procedure.

\subsection{Structure-Only Initialization via a Differentiable MD Engine}\label{sec::methods::torch-simulator}

Our first proposed solution to the SOI problem, introduced above, is to employ a custom-built differentiable MD simulator to generate the necessary initial dynamical context for the GNN simulator when trajectories are initialized from a single system configuration (see  Figure \ref{fig::diagram}). To this end, we implemented an MD engine entirely in PyTorch,\cite{Paszke2019} specifically tailored for uniaxial compression of 2D disordered elastic networks in the NPT ensemble. Similar to conventional MD engines, it uses a half-step velocity Verlet integrator coupled with a barostat. This custom-built MD engine is used to generate a short bootstrap trajectory starting from a static input configuration. The required length of the transient trajectory $l$ is determined by the required ML model history $h$ and the expected ML simulator step size $d_{cg}$, which may correspond to 10-10000 standard MD steps:
\begin{eqnarray}
    l = h \cdot d_{cg}
\end{eqnarray}
Hence, for an $h=3$ model with $d_{cg} =200$ MD steps, the default configuration we adopted, 4 static snapshots are extracted from bootstrap trajectory with $l=600$ MD steps. For full description and implementation details, see Section IV in the Supplementary Information.

\subsection{Structure-Only Initialization via Simulator Cascade}\label{sec::methods::cascade}

As an alternative solution to the SOI problem, we propose a purely data-driven simulator cascade approach (Figure \ref{fig::diagram}). The core idea is to distribute the challenge of predicting subsequent trajectory steps in the absence of sufficient dynamical input across a sequence of independent, specialized GNN simulators. This cascade serves as a transient phase, predicting the initial, more challenging steps from a static configuration to construct the required dynamical context before the primary simulator takes over. In this setup, each model, except for the final, main simulator, is tasked with predicting a single, specific timestep $t$. This specialization improves prediction accuracy by tailoring each model to the limited input information available at its respective stage.

The simulator cascade consists of a sequence of models with increasing input history requirements ($h=0,1,2, \dots $). For the base model ($h=0$), the node features of the input graph $G^{h=0}$ contain only the network's bead positions $\textbf{r}$. For the subsequent models ($h \ge 1$), bead velocity is added to the input graph node features, calculated as $\textbf{v}^t = \textbf{r}^t - \textbf{r}^{t-1}$, such that each subsequent model utilizes a larger temporal context window $(\textbf{v}^t, \textbf{v}^{t-1}, \dots, \textbf{v}^{t-h+1})$. During the initial stages of a rollout, this enables the cascade to progressively construct the temporal context required by subsequent models, thereby circumventing the SOI problem that arises when starting from a single structure at $t=0$. For details on the simulator cascade training procedure see Section \ref{sec::comp::cascade_training}.

\subsection{Inference-time Physics-based Optimization}\label{sec::methods:ITPO}

To improve out-of-distribution generalization and overall rollout stability, we introduce Inference-Time Physics-Based Optimization (ITPO). This approach refines each GNN prediction by enforcing physics-based constraints, treating the simulator output as a high-quality initial approximation that is subsequently optimized at every step to ensure consistency with the governing physical constraints. Specifically, the simulator-predicted accelerations are optimized via gradient descent against a composite loss function consisting of an anchor term $\mathcal{L}_{\text{anchor}}$ and a weighted physics term $\mathcal{L}_{\text{physics}}$:
\begin{eqnarray}
    \mathcal{L}_{\text{total}} = \mathcal{L}_{\text{anchor}} + \alpha \mathcal{L}_{\text{physics}},
\end{eqnarray}
where $\alpha$ is an empirically determined weighting coefficient, estimated from the simulator’s original training dataset. The anchor term $\mathcal{L}_{\text{anchor}}$ captures the compression dynamics learned by the GNN during training. In contrast, $\mathcal{L}_{\text{physics}}$ serves as a corrective term, steering the initial GNN prediction toward a physically consistent state by enforcing domain-specific constraints. This framework is general and can be applied to a broad class of dynamical systems, provided the relevant physical constraints are available. 

The proposed Inference-time Physics-based Optimization (ITPO) strategy strictly decouples the physical constraints from the training loop. There are several examples of analogous optimization strategies used in the literature in contexts unrelated to MD. While some such approaches rely on inference-time fine-tuning of the neural network weights,\cite{Li2024Physics-InformedEquations} ITPO aligns with a second paradigm that freezes the model weights and optimizes the physical state directly. Examples of this approach include particle interactions,\cite{Rubanova2021} rigid and soft body dynamics,\cite{Yang2020LearningProjections} inverse PDE problems,\cite{Zhao2022LearningNetworks} as well as fluid and continuous structural simulations.\cite{Rochman2025EnforcingProjections, Iftakher2026Physics-informedConstraints} 

\subsubsection*{Physical constraints}
For the uniaxial compression of DENs, we construct $\mathcal{L}_{\text{physics}}$ from three explicit physical constraints: a barostat term $\mathcal{L}_{\text{press}}$, the system's potential energy $\mathcal{L}_{\text{U}}$, and the mean-squared per-particle force $\mathcal{L}_{\text{force}}$. 

\textit{Barostat Term:} implementing the pressure-based constraint $\mathcal{L}_{\text{press}}$ requires a differentiable method to calculate and adjust the periodic box boundaries, i.e., a custom GNN simulator-based barostat. Here, we build on foundational approaches from the literature.\cite{Andersen1980, Parrinello1981, Feller1995} At each rollout timestep $t$, given the constant engineering strain applied along the $x$ direction, the algorithm computes the system's instantaneous transverse internal pressure $P_y$, which is composed of a kinetic ($P_y^K$) and a virial ($P_y^V$) component:
\begin{eqnarray}\label{eq::bar1}
    P_y = P_y^K + P_y^V = \frac{k_B NT}{A} + \frac{1}{A} \sum_{i < j} f_{ij, y} \; r_{ij, y}
\end{eqnarray}
where $A$ denotes the simulation box area, $f_{ij, y}$ is the $y$-component of the harmonic force corresponding to a bond between nodes $i$ and $j$, and $r_{ij, y}$ is the $y$-component of their relative distance vector. The deviation of $P_y$ from the prescribed target external pressure $P_t$ defines a thermodynamic driving force $F_d$. As is customary, we introduce a frictional damping coefficient $\gamma$ to mitigate numerical oscillations during the coarse-grained rollouts: 
\begin{eqnarray}\label{eq::bar2}
    F_{\text{total}} = (P_y - P_t) \cdot L_x - \gamma \; v^t_y
\end{eqnarray}
where the resulting box acceleration $a^t_y = F_{\text{total}}/W_y$ (where $W_y$ is the so called system piston mass) is used to update the box velocity $v^t_y$ and its transverse dimension $L_y$ through:
\begin{eqnarray}\label{eq::bar3}
    L^{t+1}_y = L_y^t \cdot \exp(v^{t}_y \Delta t)
\end{eqnarray}
Notably, since a single coarse-grained GNN simulator timestep can correspond to anywhere between 10 and 10,000 MD timesteps, both the damping coefficient $\gamma$ and piston mass $W_y$ must be adjusted accordingly to remain consistent with the chosen timestep. See the Supporting Information for more details.

\textit{Potential energy term:} Motivated by the observation that non-physical trajectories tend to produce a substantial increase in the system’s potential energy, defined as the sum over all bond energies, $E(\textbf{r}_{ij})$ (Section \ref{sec::methods:DENs}):
\begin{equation}
    \mathcal{L}_U = \sum_{i<j} E(\textbf{r}_{ij})
\end{equation}
we impose an upper bound on this term and effectively minimize it.

\textit{Per-Particle Net Force Term:} we found it beneficial to constrain the net forces acting on the particles, as non-physical solutions tend to increase them substantially beyond the values observed in ground-truth simulations. To penalize these forces along the $y$-axis, the per-particle net force constraint is defined as follows:
\begin{eqnarray}
    \mathcal{L}_{\text{force}} = \frac{1}{N} \sum_{i=1}^N (F_{i, y})^2
\end{eqnarray}
where $F_{i, y}$ denotes the $y$-component of the net force acting on the $i$-th particle: 
\begin{eqnarray}
    \mathbf{F}_i = \sum_{j \in \mathcal{N}(i)} -k_i (l - l_0) \frac{\mathbf{r_{ij}}}{l}
\end{eqnarray}
where $k_i$ is the harmonic bond stiffness and $l$ and $l_0$ denote the current and rest bond lengths, respectively.

\subsubsection*{Optimization procedure}

During rollouts, the ITPO loop is applied following each forward pass of the trained GNN simulator. By casting the predicted accelerations as a learnable parameter tensor, $\textbf{a}_{\text{refined}}$, we can iteratively optimize the state using the Adam optimizer.\cite{Kingma2017Adam:Optimization} The anchor constraint $\mathcal{L}_{\text{anchor}}$ penalizes the mean squared deviation between the optimized accelerations $\textbf{a}_{\text{refined}}$ and the initial predictions of the frozen GNN, $\textbf{a}_{\text{nn}}$:
\begin{eqnarray}
    \mathcal{L}_{\text{anchor}} = \frac{1}{N} \sum_{i=1} ^ N (\mathbf{a}_{\text{refined}} - \mathbf{a}_{\text{nn}})^2
\end{eqnarray}

\section{Results and Discussion}\label{sec::discussion}

\begin{figure}[htbp]
    \centering
    \includegraphics{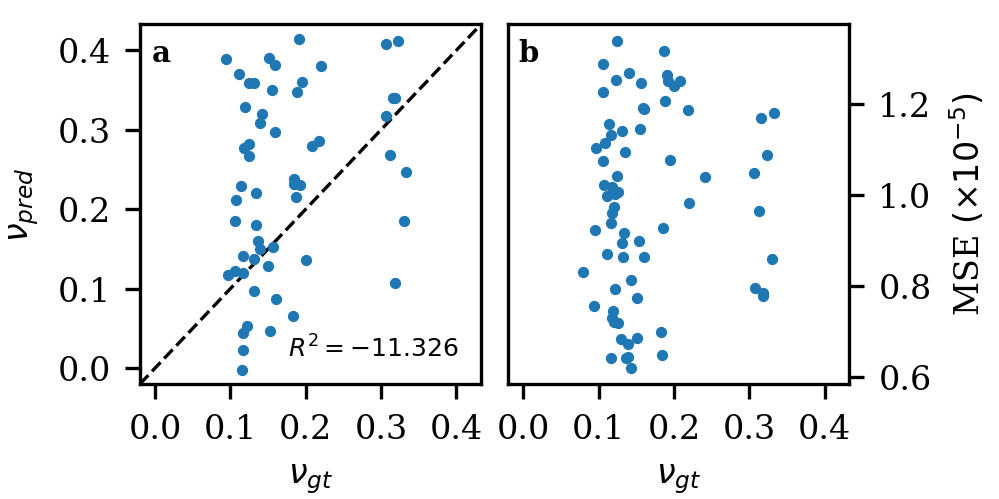}
    \caption{Performance of a position-only GNN simulator after a 10-step rollout. (a) Parity plot of predicted versus ground truth Poisson's ratio $\nu$. (b) Mean-squared position error as a function of ground truth $\nu$.}
    \label{fig::position_model}
\end{figure}

As discussed in Section \ref{sec::intro}, utilizing GNN simulators within inverse design and optimization pipelines requires addressing the SOI problem. A natural baseline is a GNN simulator trained with zero temporal history $h=0$, taking only a static configuration as input without access to particle velocities or accelerations. However, such models typically exhibit poor performance. Figure \ref{fig::position_model} shows that a position-only GNN simulator fails to sustain even a simple 10-step rollout, as evidenced by the relatively large position MSE and an inability to accurately predict macroscopic system properties such as Poisson's ratio $\nu$. This failure highlights the need for methodological advances that enable stable and accurate trajectory prediction from a single static configuration, which is an essential requirement for integrating ML-based simulators into autonomous material optimization workflows.

\begin{figure}[htbp]
    \centering
    \includegraphics{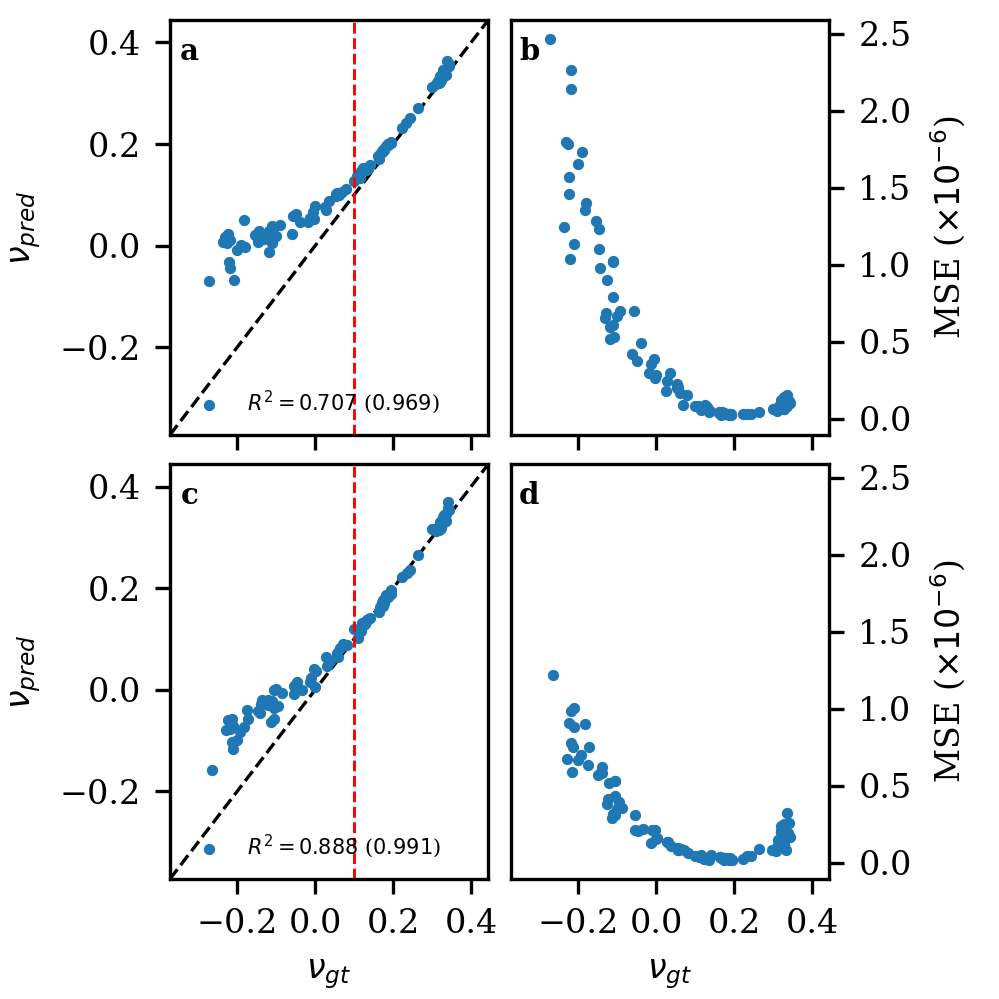}
    \caption{Performance of structure-only initialization methodologies over 50-step rollouts. (a, b) MD-bootstrapped GNN simulator. (c, d) Simulator cascade. Panels (a) and (c) show parity plots of predicted $\nu_\text{pred}$ versus ground truth $\nu_\text{gt}$ Poisson's ratios. Panels (b) and (d) show the mean-squared position error ($\times 10^{-6}$) as a function of $\nu_{\text{gt}}$. Dashed black and red lines represent the identity $\nu_{\text{gt}} = \nu_{\text{pred}}$ and the training data cut-off at $\nu =0.1$, respectively. All models were trained using multi-step supervision.}
    \label{fig::acc_ft}
\end{figure}

As a first potential solution to the problem of structure-only trajectory initialization illustrated in Figure \ref{fig::position_model}, we utilize a short bootstrap trajectory generated by a minimal custom-built MD engine. The performance of this approach is shown in Figure \ref{fig::acc_ft}a and b. To explore the OOD generalization capabilities of the GNN simulator, the training dataset was restricted to Poisson's ratios $\nu \geq 0.1$. As previously reported,\cite{Salman2025EvaluatingNetworks} the compression dynamics of non-auxetic ($\nu \geq 0$) elastic networks are notably less complex than those of auxetic networks ($\nu < 0$) (see Figures S5 and S6 in Supplementary Information), making this a particularly good case study for OOD performance. In contrast to the position-only baseline, our bootstrapped GNN simulator is highly accurate on "in-distribution" validation data ($\nu \geq 0.1$), reaching $R^2 \approx 0.97$ (Figure \ref{fig::acc_ft}a). Furthermore, it demonstrates significant OOD generalization, maintaining an $R^2$ of approximately $0.71$ across the full $\nu$-range.

As GNN simulators, once trained, can offer a computational advantage over both conventional differentiable MD engines and highly optimized MD engines such as LAMMPS through the coarse-graining of the dynamics, the bootstrapped GNN simulator remains effective. Although several general-purpose differentiable MD frameworks have recently emerged,\cite{Schoenholz2021, Doerr2021TorchMD:Simulations, Ple2024FeNNol:Potentials, Greener2024DifferentiableProteins, Christiansen2025FastSimulations, Cohen2025TorchSim:PyTorch} their direct application to inverse design is often hindered by vanishing or noisy gradients over long trajectories. While the computational cost of generating a short bootstrap trajectory is relatively small, this step can introduce minor overhead depending on the required temporal history of the model $h$. In addition, the lack of a universal differentiable PyTorch-based MD framework may necessitate further implementation effort when the target physical system is not supported by existing simulators.

Since bootstrapping a GNN simulator with a transient trajectory generated by a custom differentiable MD engine involves certain trade-offs, we propose a fully data-driven alternative as well: the simulator cascade. In this strategy, the prediction of the initial transient trajectory is handled by a sequence of specialized GNN simulators applied sequentially during the early stages of the rollout (see Section \ref{sec::methods::cascade} for more details). As can be seen in the parity plot (Fig. \ref{fig::acc_ft}c), the simulator cascade demonstrates superior OOD generalization, achieving $R^2 \approx 0.89$ on data with $\nu$ down to $-0.3$ and an $R^2 \approx 0.99$ for in-distribution non-auxetic data. Interestingly, a distinct difference between the two methods can be observed in the position MSE plots (Fig. \ref{fig::acc_ft}b and d). While the bootstrapped GNN simulator maintains a relatively flat MSE plateau for test data with $\nu \geq 0.1$, the simulator cascade approach features a clearly defined minimum around $\nu \approx0.2$, with the error increasing in both positive and negative directions. 

As a simpler alternative for addressing the SOI problem, we also explored a “padded input” GNN simulator. The architecture is identical to the standard model, with the key difference lying in the input representation. In contrast to the strategies above, this model initializes the missing dynamical and historical context at the start of a trajectory, i.e., velocities from the current and previous timesteps ($v^t, v^{t-1}, \dots$), to zero. Additionally, input node features $\textbf{x}$ for each $G^t$ include the current timestep $t$ and positions $\textbf{r}^t$. As the required temporal window is progressively constructed through the autoregressive rollout, the corresponding velocity features are populated up to a predefined history length $h$. While this approach is methodologically and architecturally simpler than the methods proposed above, its OOD generalization performance is inferior, as shown in Figure S1 of the Supplementary Information.

\subsection*{Inference-time Physics-based Optimization}\label{sec::discussion::ITPO}


While GNN simulators can exhibit significant potential for OOD generalization by learning the underlying local multi-body dynamics of a system, \cite{Salman2025EvaluatingNetworks} this capability is naturally limited. The success of a purely data-driven GNN simulator fundamentally relies on learned statistical correlations. Consequently, its generalization capacity is inherently limited by the degree to which the physical regime of the unseen data diverges from the training distribution. When applied to scenarios with distinct dynamics, such as auxetic materials with negative Poisson’s ratios ($\nu < 0.0$), the GNN simulator's prediction accuracy degrades. This becomes especially important in the context of inverse design, where the explicit goal is to discover novel engineering solutions that often lie well outside the training data domain. 

To extend the GNN simulator's ability to predict trajectories and subsequent emergent properties beyond the training distribution, we propose here Inference-time Physics-based Optimization (ITPO). As described in Section \ref{sec::methods:ITPO}, the core idea of this approach is to constrain the neural network's predictions to a domain of solutions that is consistent with the underlying physics of the system. Importantly, we evaluate OOD generalization in the challenging setting of auxetic networks characterized by negative Poisson's ratios, which where strictly excluded from the training set. Given that the dynamics of auxetic networks differ from, and are often more complex than, those of non-auxetic systems, achieving accurate generalization in this regime represents a robust demonstration of the model's extrapolative capabilities.

\begin{figure}[htbp]
    \centering
    \includegraphics{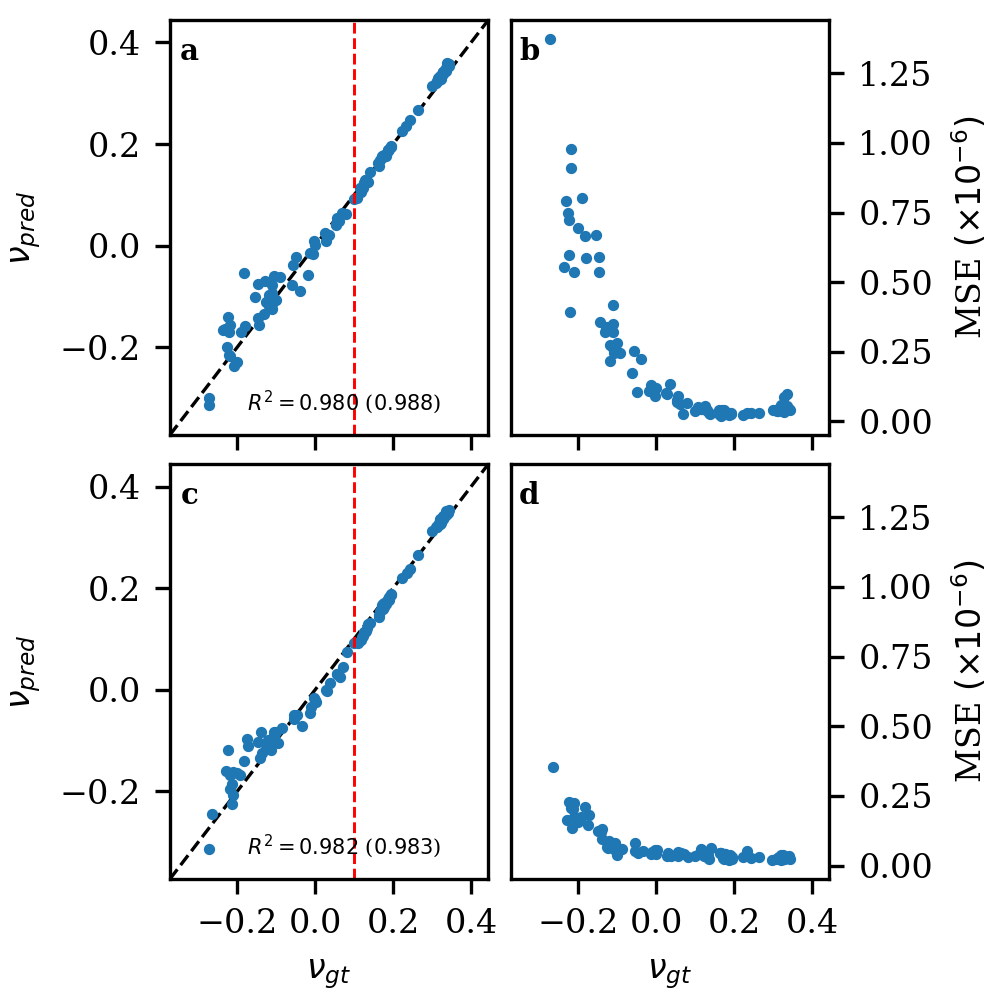}
    \caption{Performance of structure-only initialization methodologies over 50-step rollouts using ITPO. (a, b) MD-bootstrapped GNN simulator. (c, d) Simulator cascade. Panels (a) and (c) show parity plots of predicted $\nu_\text{pred}$ versus ground truth $\nu_\text{gt}$ Poisson's ratios. Panels (b) and (d) show the mean-squared position error ($\times 10^{-6}$) as a function of $\nu_{\text{gt}}$. Dashed black and red lines represent the identity $\nu_{\text{gt}} = \nu_{\text{pred}}$ and the training data cut-off at $\nu =0.1$, respectively.}
    \label{fig::ITPO}
\end{figure}

\begin{figure}[htbp]
    \centering
    \includegraphics{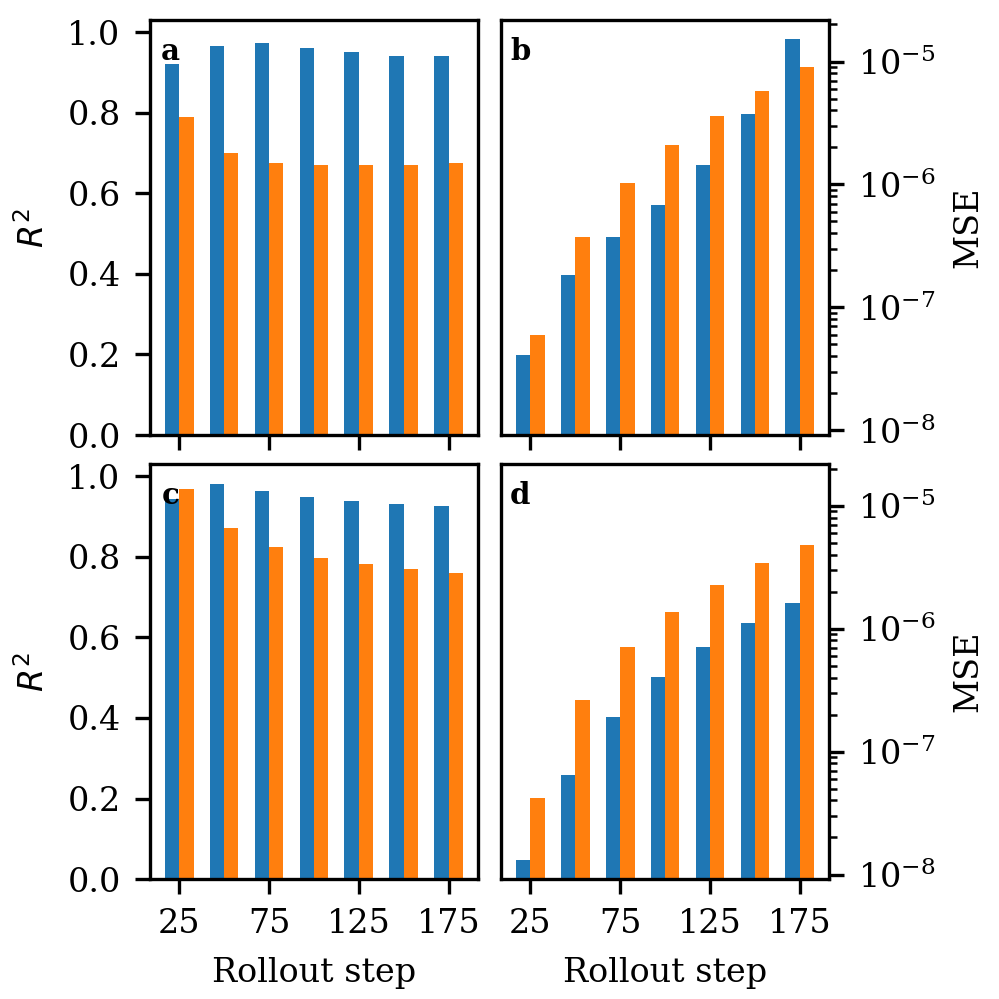}
    \caption{Performance of bootstrapped GNN simulator (a, b) and simulator cascade (c, d) as a function of rollout length. (a, c) Coefficient of determination $R^2$ for Poisson's ratio $\nu$, (b, d) average mean-squared position error. Blue and orange bars denote results obtained with and without using ITPO, respectively.}
    \label{fig::stability}
\end{figure}

As illustrated in Figure \ref{fig::ITPO}, ITPO leads to a substantial improvement in OOD generalization for both the MD-bootstrapped GNN simulator and the simulator cascade. In both cases ITPO enables the corresponding models to achieve an $R^2 \approx 0.98$ across the full $\nu$-range of validation data, which constitutes near-perfect OOD generalization. Importantly, the in-distribution $R^2$ is only marginally higher for the MD-bootstrapped GNN simulator (Fig. \ref{fig::ITPO}a), while for the cascade the metrics are virtually identical (Fig. \ref{fig::ITPO}c). Furthermore, the position MSE plots (Fig. \ref{fig::ITPO}b and d) show a consistent error plateau down to $\nu \approx 0.0$, with no error increase toward positive $\nu$, in contrast to the behavior previously observed for the simulator cascade (Fig. \ref{fig::acc_ft}d). Notably, ITPO also leads to significant improvement in rollout stability, with model performance remaining consistently high across rollouts lengths up to 200 timesteps (equivalent to 40000 MD timesteps). Figure \ref{fig::stability} compares standard and ITPO rollouts, demonstrating that ITPO yields higher $R^2$ and lower position MSE for both the MD-bootstrapped GNN simulator and the simulator cascade.

\section{Conclusions}\label{sec::conclusion}

In this work, we address the structure-only initialization (SOI) problem, a key limitation in the use of GNN-based MD simulators for inverse design and optimization, while also improving their out-of-distribution (OOD) generalization. Progress on both fronts is essential for gradient-based optimization pipelines, where a simulator must reliably predict the dynamics of novel candidate structures starting from a single static configuration that typically is not found in the available training data domain.

Using the uniaxial compression of disordered elastic networks (DENs) as a case study, we introduced two distinct strategies to overcome this "cold start" problem. Both the differentiable MD bootstrapping engine and the data-driven simulator cascade effectively address the SOI problem, enabling stable rollouts starting from purely static configurations. These tools enable ML-based dynamics to be coupled directly with structural optimizers while preserving end-to-end differentiability. Notably, the padded-input simulator, despite its overall lower performance relative to the two primary approaches, also resolves the SOI problem and may be advantageous in certain settings due to its simpler architecture. 

Beyond these proposed approaches, we additionally demonstrated the improvements derived from Inference-time Physics-based Optimization (ITPO). ITPO serves as a corrective mechanism that enforces physical consistency during inference. Specifically, models trained exclusively on high Poisson’s ratio data ($\nu \geq 0.1$) were shown to accurately reproduce the macroscopically distinct and substantially more complex dynamics of highly auxetic elastic networks down to $\nu = -0.3$. It is worth noting that while ITPO is highly effective at improving both OOD generalization and prediction accuracy, it introduces additional computational overhead. Future work exploring mitigation strategies, such as limiting the number of optimization steps or applying ITPO periodically rather than at every rollout step, may help reduce this cost.

Furthermore, we introduced a GNN-based barostat that dynamically regulates periodic system dimensions and the resulting pressure. This proved to be essential for capturing macroscopic system properties, e.g., Poisson's ratio $\nu$, and generalizing beyond the training data distribution. Complementing the proposed methodologies, we employed a Multi-Step Training (MST) regime throughout this study. MST improves rollout stability and reduces error accumulation by exposing the model to its own state drift during training. 

We find that, together, the proposed strategies enable structure-only rollout initialization, significantly suppress error accumulation, and improve OOD generalization, extending prediction capabilities to regimes that exhibit distinct and more complex physical behavior than those present in the training data. Ultimately, these results indicate that, when properly initialized and constrained by physical priors, GNN-based simulators can serve as efficient and generalizable tools for material discovery and structural optimization.

\section{Computational Details}\label{sec::comp}

\subsection{Disordered elastic networks and dataset generation}\label{sec::methods:DENs}
 
Disordered elastic networks (DENs) are highly tunable systems, consisting of point-sized beads connected by harmonic bonds, with an energy in the form:
\begin{eqnarray}
    E(\textbf{r}_{ij}) = K_b(l-l_0)^2.
\end{eqnarray}
Initial DEN configurations were generated in a two-step process\cite{Rocks2017} using a standard jamming algorithm.\cite{Liu2010}

Training data was generated using the LAMMPS molecular dynamics package.\cite{Thompson2022} Each dataset entry corresponds to a compression trajectory, a sequence of system configurations sampled from a ground truth simulation at regular timesteps $t$, beginning from the initial timestep $t = 0$. By default, generated disordered elastic networks are characterized by relatively high Poisson's ratio $\nu \approx 0.3$. To generate test and validation data with $\nu < 0$, two different optimization procedures were employed. The first procedure uses a constrained gradient descent algorithm \cite{Shen2024}, which optimizes the node positions $\textbf{r}$ to achieve a lower $\nu$, while the network topology satisfies angle and distances constraints. The second optimization procedure, which leads to networks very different in form and behavior, relies on a sequential pruning strategy,\cite{Reid2018} where at every optimization step a single bond, which leads to the lowest change in shear modulus $\Delta G$, is pruned from the network. In this work we utilize a custom strategy inspired by the pruning algorithm. By utilizing an in-house implemented differentiable MD engine, we directly optimize network bond stiffnesses $k_i$ to reduce $\nu$. See Section I of the Supplementary Information for more details on DENs, their optimization and dataset generation procedures. Examples of disordered elastic network topologies obtained via these optimization methods are shown in Fig. \ref{fig::dens}b and c.

\subsection{Data representation}\label{sec::comp::data_repr}

The system configuration at each timestep $t$ is encoded into a graph $G^t$, consisting of a set of nodes $V$ of size $N$ and a set of edges $E$ of size $M$. Each node $i$ in $V$ corresponds to a bead in the disordered elastic network and is assigned a corresponding feature vector $\textbf{x}_i$. These vectors typically contain the current timestep per-node positions $\textbf{r}^t_i$ or per-node velocities $\textbf{v}^t_i$. Assuming a unit timestep ($\Delta t = 1$), the velocities are calculated as the finite difference between the current and previous positions: $\textbf{r}^t_i - \textbf{r}_i^{t-1}$. Depending on the model's temporal history $h$, a node feature vector aggregates an equal number of current and past velocities ($\textbf{v}^t_i, \textbf{v}^{t-1}_i, \dots, \textbf{v}^{t-h+1}_i$). Each edge $e_{ij}$ between nodes $i$ and $j$ contains a feature vector composed of the relative edge vector $\textbf{r}_i - \textbf{r}_j$, the harmonic bond length $l$, and the bond stiffness $k$, defined as the inverse of the rest length ($1/l_0$).

\subsection{GNN simulator training}\label{sec::methods::mst}

Conventionally, direct GNN simulators are trained by supervising on a single forward step, typically minimizing the error of per-particle accelerations $\textbf{a}^t_i$ or velocities $\textbf{v}^t_i$.\cite{Pfaff2021} This strategy fails to account for error accumulation during inference, as the simulator is never exposed to its own rollout errors. As a result, the model tends to overfit to one-step predictions, leading to degraded accuracy over longer rollouts. 

A common workaround is to corrupt the training inputs with artificial Gaussian noise to emulate error accumulation \cite{Sanchez-Gonzalez2020, Pfaff2021} and mitigate "oversmoothing".\cite{Godwin2022SimpleBeyond} The natural disadvantage of this strategy is the need to chose what distribution this artificial noise should follow, as well as the magnitude of that noise. Moreover, when OOD generalization is the goal, there is no guarantee that injected synthetic noise accurately reflects the model's actual error both in seen and unseen physical regimes. If the target material properties are absent from the training data, tuning an appropriate noise profile becomes effectively infeasible. 

To mitigate the challenges stated above, we devised a multi-step supervision training strategy (MST). Instead of isolated one-step predictions, the model performs a short, continuous rollout during the forward pass for every training sample. By supervising across the entire trajectory, we expose the model to its own state drift, which encourages it to actively learn "self-correction", forcing its predictions towards stable physical dynamics before the errors can compound. Similar approaches have been reported previously.\cite{Brandstetter2023MessageSolvers, Hoang2025, Tian2026ScalingNetworks}

At each step in this rollout, we compute the per-step loss $\mathcal{L}_{step}$ between the model's prediction and the corresponding ground truth. The total loss for the trajectory is then computed as the average of these individual step losses:
\begin{eqnarray}
  \mathcal{L}_{total} = \frac{1}{k}\sum_{k} \mathcal{L}_{step, k}
\end{eqnarray}
where $k$ is a rollout length, which is treated as a hyperparameter and is tuned for each dataset/model combination.

Crucially, before using the model's current timestep prediction as the input for the subsequent timestep, we detach the predicted state from the computational graph. This allows the model to learn how to make stable predictions from slightly erroneous input configurations, effectively teaching it "self-correction" without the computational penalty of back-propagating through the entire rollout.

To illustrate the improvement from the multi-step training strategy over the conventional one-step supervision training, we compared the performance of otherwise identical $h=3$ GNN simulators and simulator cascades (Figures S3 and S4 in Supplementary Information).

\subsection{Simulator cascade training}\label{sec::comp::cascade_training}

The training process for the simulator cascade follows a two-stage procedure: sequential progressive training followed by an end-to-end refinement run.

In the first stage, each model in the cascade is trained sequentially and independently. When training a target model $k$, all preceding models ($0$ to $k-1$) are instantiated and their weights are frozen. To generate the training inputs for the $k$-th model, the frozen $k-1$ models perform a short rollout starting from a ground-truth static configuration. The $k$-th model then receives this generated trajectory as its input history and is trained to make a single-step prediction to match the ground-truth target. This approach exposes the target model to the accumulated prediction errors of the previous models, teaching it to inherently correct for deviations early in the trajectory.

The second stage of the training procedure is an end-to-end refinement run, during which all models in the cascade are unfrozen and their weights are optimized together. Crucially, the gradients flow across the entire unrolled cascade of different models. The detailed training procedure is available in Section VA of the Supplementary Information.

\subsection{Simulator-coupled barostat}

To quantify the macroscopic behavior of the DENs, such as their Poisson’s ratio $\nu$, it is necessary to accurately measure their dimensional changes during compression. While the compression axis $x$ follows a constant engineering strain rate, the transverse axis $y$ expands and contracts according to the network’s intrinsic Poisson’s ratio . To accurately capture this lateral strain, which is a necessary feature for out-of-distribution (OOD) generalization, we utilize the differentiable GNN-simulator-based barostat (Eqs. \ref{eq::bar1} - \ref{eq::bar3}). This dynamic box adjustment procedure determines the 2D periodic box dimensions that correspond to the system's target pressure, particle positions, and forces (Eq. \ref{eq::bar1}). It also serves as a foundation that enables both the multi-step supervision (MST) and Inference-time physics optimization (ITPO) strategies (detailed in Sections \ref{sec::methods::mst} and \ref{sec::methods:ITPO}, respectively).

\begin{figure}[htbp]
    \centering
    \includegraphics{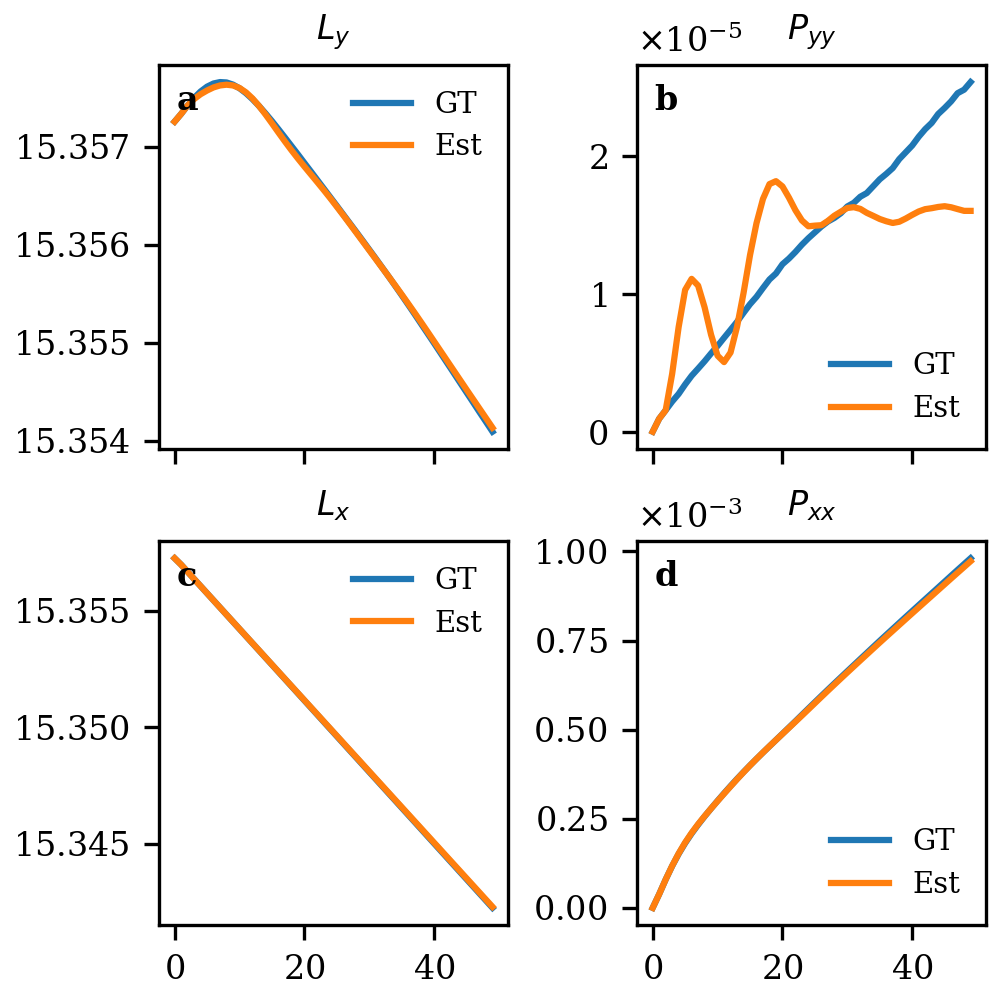}
    \caption{Comparison between dynamically calculated (orange) and ground truth (blue) box dimensions $L_y$ (a) and $L_x$ (c), as well as the internal virial pressure along $y$ (b) and $x$ (d) axes.}
    \label{fig::barostat_box_behaviour}
\end{figure}

We benchmark the accuracy of the simulator barostat and the chosen piston mass $W_y$ and frictional damping coefficient $\gamma$ within the context of coarse-grained dynamics by replicating a model rollout using ground truth compression data. More specifically, the GT trajectories are coarse-grained (1 step equals 200 MD steps) and the GT bead positions are used instead of predicted ones. Figure \ref{fig::barostat_box_behaviour} illustrates how this procedure manages to replicate the behavior of the 2D periodic box during the compression trajectory. While $L_x$ changes linearly due to the constant engineering strain rate (Fig. \ref{fig::barostat_box_behaviour}c), $L_y$ has a more interesting behavior, showing a brief period of expansion followed by a linear decline (Fig. \ref{fig::barostat_box_behaviour})a. We observed this specific $L_y$ behavior to be common among virtually every auxetic DEN, generated either via global node optimization or via bond stiffness optimization (see Section IC in the Supplementary Information). Additionally, the simulator-coupled barostat successfully manages to keep the pressure $P_{yy}$ close to the ground truth (Fig. \ref{fig::barostat_box_behaviour}b). A significant fluctuation can be observed, which can be attributed to coarse-grained nature of dynamics, i.e., since $L_y$ is only allowed to relax every 200 MD steps, the pressure difference has more time to build up, causing the observed spikes.

\begin{acknowledgments}
The authors acknowledge support from the Israel Science Foundation (ISF) under grant number 1181/24.
\end{acknowledgments}

\bibliography{papers} 

\end{document}



\title{Supporting Information}

\author{S. A. Shteingolts}
\author{Salman N. Salman}%
\author{Dan Mendels}
\email{danmendels@technion.ac.il}
\affiliation{The Wolfson Department of Chemical Engineering, Technion – Israel Institute of Technology, Haifa, 32000, Israel}

\date{\today}

\maketitle

\section{Disordered Elastic Networks and Data generation}

\subsection{Elastic network generation}
Disordered elastic networks (DENs) were generated in a two-step process.\cite{Rocks2017} First, $N$ soft disks of 4 different types and diameters were randomly placed in equal proportions in a periodic simulation box and allowed to relax to a local energy minimum using a standard jamming algorithm.\cite{Liu2010} The frictional force between two disks $i$ and $j$, characterized by their respective radii $R_i$ and $R_j$, within a contact distance $d=R_i + R_j$ is defined as:
\begin{eqnarray}
F=(k_n\delta \textbf{n}_{ij} - m_{eff}\gamma_n \textbf{v}_n) - (k_t \Delta s_t + m_{eff} \gamma_t \textbf{v}_t),
\end{eqnarray}

where $\delta$ is an overlap distance, $\textbf{n}_{ij}$ is the unit vector along the line connecting the centers of two disks, $k_n$ and $k_t$ are elastic constants for normal and tangential contact, respectively, $\gamma_n$ and $\gamma_t$ are viscoelastic constants for normal and tangential contact, $m_{eff} = \frac{M_i M_j}{M_i + M_j}$ is an effective mass of two disks $i$ and $j$ with individual masses $M_i$ and $M_j$, $\textbf{v}_n$ and $\textbf{v}_t$ are the normal and tangential components of the relative velocity of the two disks, and $\Delta s_t$ is the tangential displacement vector between the two disks. In our case, $k_t$, $\gamma_n$ and $\gamma_t$ were set to zero, while $k_n$ was set to 1.0, so the resulting force between the two disks is simply expressed as follows:
\begin{eqnarray}
F=\delta \textbf{n}_{ij}
\end{eqnarray}

During the energy minimization, the system was slowly cooled from $T_{start}=5\times10^{-6}$ to $T_{end}=1\times10^{-6}$.
After that, the beads were placed at the center of each disk and harmonic bonds were created between beads that were in contact  $\textbf{r}_{ij} \leq d$. Harmonic bonds were assigned an elastic energy in the form:
\begin{eqnarray}\label{eq::harmonic_energy}
E(r_{ij}) = k (l - l^{0})^2, 
\end{eqnarray}
where $k$ is the bond stiffness with units $energy/distance$, $l^0$ is the bond rest length, and $l$ is the distance between the two beads. For each bond, the value of $k$ was set to $1/l^0$. The resulting network configuration was then scanned for "dangling" beads, i.e., those with less than 3 connections to their neighbors, which were removed. The scanning-removal procedure was applied iteratively until no such beads were left.

\subsection{Uniaxial compression simulation}
Compression simulations of the 2D DENs were carried out using the LAMMPS \cite{Thompson2022} molecular dynamics simulator. The systems were simulated within the NPT ensemble with anisotropic pressure control (applied to the simulation box in the direction perpendicular to compression), with constant temperature $T=1\times 10^{-10}$ , $k_B = 1$ , damping parameter of $7.1 \times 10^{-3}$ and time step $\tau = 7.13 \times 10^{-6}$ controlled by the Langevin thermostat.\cite{Schneider1978} The compression was simulated using the \texttt{fix deform} command of LAMMPS with a constant engineering strain rate $s_r = 1\times10^{-8}$ and the maximum strain of 3\% along the compressed axis. Constant zero-pressure along the non-compressed axis was applied on the simulation box perpendicular to the direction of compression using the Parrinello-Rahman barostat.\cite{Parrinello1981}

\subsection{Poisson ratio optimization}

\subsubsection{Node displacement-based optimization}\label{sec::opt::global_node}

For low-Poisson's ratio, including auxetic, network generation the global node optimization strategy was employed.\cite{Shen2024} The strategy is based on stochastic gradient descent and aims to minimize Poisson's ratio $\nu$ by altering individual node positions. At each optimization epoch $n$, the coordinate of each node $i$, $\textbf{r}_i$ is displaced by $5 \times 10^{-4}$ in an axis $\eta \in \{x,y\}$. After that, bond lengths and stiffnesses are recalculated such that the system is in its equilibrium configuration and the gradient is computed as follows: 
\begin{eqnarray}
\frac{\partial(\nu + \mathcal{L}_1 + \mathcal{L}_2)}{\partial r_{i\eta}},
\end{eqnarray}
where $r_{i\eta}$ is the displacement of node $i$ in the axis $\eta$, while $\mathcal{L}_1$ and $\mathcal{L}_2$ are the internode distance and angle constraints, respectively, which are defined as follows:
\begin{eqnarray}
\mathcal{L}_1 &=& 0.1 \sum_{i, j} H(r_{min} - r_{ij})(r_{ij} - r_{min})^2, \\
\mathcal{L}_2 &=& 0.01 \sum_{j} H(\theta_{min} - \theta_{j})(\theta_{j} - \theta_{min})^6, 
\end{eqnarray}
where $H$ is the Heaviside step function, while $r_{min}=0.3r^0$ and $\theta_{min}=15^{\circ}$ are the empirically chosen minimal values for bond distances and angles, respectively. Both $\mathcal{L}_1$ and $\mathcal{L}_2$ serve to constrain the optimization space such that the networks beads don't overlap and bonds do not intersect. To prevent structure from changing too aggressively and make the optimization process more stable, gradients were clipped to 0.01 by absolute value. Finally, the initial node positions are updated using the $N\times2$ gradient matrix obtained at the previous step:
\begin{eqnarray}
r_{i\eta}^{n+1} = r_{i\eta}^{n} + \lambda \frac{\partial(\nu + \mathcal{L}_1 + \mathcal{L}_2)}{\partial r_{i\eta}},
\end{eqnarray}
where $\lambda$ is the learning rate. The process is repeated iteratively until either \textit{a}) the desired target value for $\nu$ is reached or \textit{b}) the number of iterations exceeds the predefined maximum limit.

\subsubsection{Stiffness-based optimization}\label{sec::opt::stiff}

Using the minimal differentiable PyTorch-based MD engine described in Section \ref{sec::torch-sim}, we have investigated two distinct optimization schemes, referred to as \textbf{continuous} and \textbf{discrete}. In both cases, the learnable parameters in the optimization procedure are the adjustments to the individual bond stiffnesses $k_i$, denoted as $\Delta k_{i}$ for a directed edge $i$ representing a harmonic bond. 

In the \textbf{continuous optimization} scheme, initial bond stiffnesses $k^0$ are adjusted continuously:
\begin{eqnarray}
k_{i}^{\prime} = k_{i}^0 \left(1 + C \cdot \tanh(\Delta k_{i})\right),
\end{eqnarray}
where $C$ is a predefined constant limiting the maximum allowed change in stiffness.

Conversely, the \textbf{discrete optimization} scheme, inspired by a similar procedure published elsewhere,\cite{Reid2018} limits the allowed bond stiffness change, which closely resembles discrete bond pruning. This is achieved using an edge mask $g_i$:
\begin{eqnarray}
g_{ij} = \sigma(\beta \Delta k_{ij}^{\text{sym}}), \\
k_{i}^{\prime} = k_{i}^0 g_{i}, 
\end{eqnarray}
where $\sigma$ is the sigmoid function and $\beta$ is a steepness hyperparameter.

At each optimization epoch, a DEN with updated stiffnesses $k^{\prime}$ is compressed for a fixed number of MD steps. The Poisson's ratio $\nu$ of the network is calculated directly from the change in the periodic box dimensions between the initial ($t=0$) and final ($t=t_{\text{end}}$) simulation steps:
\begin{eqnarray}
\nu = -\frac{L_y(t_{\text{end}}) - L_y(0)}{L_x(t_{\text{end}}) - L_x(0)}.
\end{eqnarray}

The optimization objective is to minimize the mean squared error (MSE) between the measured Poisson's ratio and the target value. The total loss function $\mathcal{L}$ is defined as:
\begin{eqnarray}
\mathcal{L} = (\nu - \nu_{\text{target}})^2 + \lambda \frac{1}{M} \sum_{ij} (1 - g_{ij}),
\end{eqnarray}
where $M$ is the total number of unique edges. The second term is a sparsity regularization term exclusive to the discrete mode and $\lambda$ is the sparsity regularization weight. For the continuous mode, $\lambda$ is set to 0. 

Optimization was performed using the Stochastic Gradient Descent (SGD) algorithm with momentum. Gradient clipping with a maximum norm of 1.0 was applied to the stiffness adjustments $\Delta k_{i}$ before each optimizer step. The procedure iterates for a predefined number of epochs or is terminated early if the simulated network successfully achieves a target Poisson's ratio $\nu_{\text{target}}$.

\section{Simulator-coupled barostat}\label{sec::barostat}

\subsection{Implementation}

At each timestep $t$, the box adjustment algorithm computes the instantaneous transverse virial $P_y^V$ and kinetic $P_y^K$ pressures directly from the  node positions predicted by the GNN simulator and pairwise forces. The kinetic term is computed using the ideal gas approximation as follows:
\begin{eqnarray}
    P_y^K = \frac{1}{2}\frac{k_B NT}{A},
\end{eqnarray}
while the virial term is computed by summing up the force contributions from all harmonic bonds in the elastic network:
\begin{eqnarray}
    P_y^V = \frac{1}{2A} \sum \textbf{f}_{ij, y} \cdot \textbf{r}_{ij, y}
\end{eqnarray}
Here $A$ is the box area, $\textbf{f}_{ij, y}$ is the $y$ component of the harmonic force $\textbf{f}_{ij}$:
\begin{eqnarray}
    \textbf{f}_{ij} = -k (\textbf{r}_{ij}-\textbf{r}_{ij, 0})
\end{eqnarray}
where  $k$ is the bond spring constant. The difference between this internal pressure $P_y$ and the target external pressure $P_t$ generates a thermodynamic driving force $F_d$. We define a total force $F_{\text{total}}$ and apply a frictional damping coefficient $\gamma$ to prevent aggressive oscillations:
\begin{eqnarray}
    F_{d} = (P_y - P_t) \cdot L_x \\
    F_{\text{total}} = F_d - \gamma \cdot \textbf{v}^t_y
\end{eqnarray}
where $\textbf{v}^t_y$ is the box velocity along the $y$ axis. Box acceleration $\textbf{a}^t_y$ can then be computed using the piston mass $W_y$:
\begin{eqnarray}
    \textbf{a}^t_y = \frac{F_{\text{total}}}{W_y}
\end{eqnarray}
Finally, the box velocity is updated and the next-step $L_y$ is calculated as follows:
\begin{eqnarray}
    &\textbf{v}^{t+1}_y = \textbf{v}^{t}_y + \textbf{a}^t_y \cdot \Delta t \\
    &L^{t+1}_y = L_y^t \cdot \exp(\textbf{v}^{t+1}_y\Delta t)
\end{eqnarray}

\subsection{Empirical parameter tuning}

Because the periodic box adjustment procedure operates on coarse-grained dynamics (a single GNN step corresponds to 200 MD timesteps), the physical constants typically used for piston mass $W_y$ and damping coefficient $\gamma$ in classical barostats are not directly applicable. Consequently, the specific values of both $\gamma$ and $W_y$ must be empirically tuned. 

To this end, we utilize the Optuna hyperparameter optimization framework. \cite{Akiba2019Optuna} The piston mass and damping coefficient are defined as functions of the number of particles $N$ and the coarse-grained time stride $\Delta t$:
\begin{eqnarray}
    W_y &=& C_{\text{coupling}} \cdot N \cdot (\Delta t)^2 \\
    \gamma &=& C_{\text{damping}} \cdot N \cdot \Delta t
\end{eqnarray}
where $C_{\text{coupling}}$ and $C_{\text{damping}}$ are the dimensionless scalar factors to be optimized. 

To find the optimal values for $C_{\text{coupling}}$ and $C_{\text{damping}}$, we minimize a cumulative objective function. The objective function calculates the discrepancy between the predicted transverse box dimension $L_y^{\text{pred}}$ and the ground-truth dimension $L_y^{\text{GT}}$ using a Huber loss. During the optimization procedure, we utilize ground-truth particle positions at each step to isolate the box adjustment from GNN prediction errors. In our testing, 200 optimization trials were enough to fully converge on optimal values. The search space for $C_{\text{coupling}}$ was defined as $[1.0, 15.0]$ and for $C_{\text{damping}}$ as $[0.5, 10.0]$. The resulting best parameters were then fixed for all subsequent rollout experiments and MST training regimes.

\section{Default GNN Simulator}\label{sec::gnn-simulator}

\subsection{Model architecture}

The model follows an encoder-processor-decoder architecture.\cite{Hamrick2018RelationalMachines, Pfaff2021} The encoder maps the input node and edge features of a given graph $G$ into latent vectors. To ensure axis-invariant learning of the temporal dynamics, the temporal history of the node features $\{\textbf{x}_i\}$ is first reshaped into separate spatial dimensions ($x$ and $y$). These are processed by an axis-shared MLP before being flattened and linearly projected into a hidden dimension of 128. Edge features $\{\textbf{e}_i\}$ are mapped directly to the 128-dimensional latent space using an independent MLP $\varepsilon^e$. 

The processor consists of $k$ identical message passing (MP) layers, each with its own set of learnable parameters. These layers are applied sequentially, with each layer operating on the output of the previous one with an addition of a residual connection. Node and edge embeddings are updated according to the following relation: 
\begin{eqnarray}
    \textbf{x}_{i}^{(k)} = \textbf{x}_{i}^{(k-1)} + \psi^{(k)} \left( \textbf{x}_{i}^{(k-1)}, \sum_{j\in\mathcal{N}(i)}\phi^{(k)}(\textbf{x}_{i}^{(k-1)}, \textbf{x}_{j}^{(k-1)},\textbf{e}_{ij}) \right), 
\end{eqnarray}
where $\psi$ and $\phi$ are learnable MLPs utilizing layer normalization. Here, $\textbf{x}_{i}^{(k-1)}$ denotes the embedding of node $i$ in the $(k-1)$th layer of the processor, and the summation is taken over all nodes $j$ which are bonded to $i$.

The output of the processor is mapped by a decoder to a two-dimensional output $\{\hat{\textbf{a}}_i\} \in \mathbb{R}^{N \times 2}$, representing the predicted node accelerations, using an additional MLP denoted $\delta^x$. All MLPs in the architecture, including $\varepsilon^x, \varepsilon^e, \psi, \phi$, and $\delta^x$, were implemented as feed-forward networks with hidden dimensions of 128 and ReLU activations.

\subsection{Implementation and training details}\label{sec::gnn_simulator::training}

The model was implemented using PyTorch Geometric 2.5.2 \cite{Fey2019} and PyTorch 2.2.2.\cite{Paszke2019} To generate the datasets, we used DENs' uniaxial compression trajectories, consisting of systems with varying Poisson's ratios $\nu \in [-0.4,0.4]$. 

Both input (i.e., node and edge features) and output data were online normalized separately during the training to zero-mean and unit-variance. Namely, at each forward step, the data were normalized with respect to all their past values seen by the model up to that point.\cite{Sanchez-Gonzalez2020, Kumar2022GNS:Modeling} To stabilize late-stage training, the normalizer statistics were subsequently frozen after a predefined epoch, typically set to 5. All learnable weights and biases of the model were randomly initialized from a normal distribution with zero mean and a standard deviation of $1/\sqrt{n}$, where $n$ is the number of weights in the layer, as we found $\mathcal{N}(0,1/\sqrt{n})$ yields better performance overall than a uniform distribution $\mathcal{U}(-1/\sqrt{n},1/\sqrt{n})$. 

Optimization was performed using the Adam optimizer with weight decay set to zero. The learning rate was decayed at each epoch using an exponential scheduler ($\gamma = 0.995$). Gradient clipping with a maximum norm of 1.0 was applied alongside gradient accumulation. The model was trained in a supervised manner to minimize the Huber loss \cite{Huber1992RobustParameter} between the predicted accelerations $\hat{\textbf{a}}$ and the normalized ground truth target $\textbf{a}$:
\begin{eqnarray}
L_{\delta}(\textbf{a}, \hat{\textbf{a}}) = 
\begin{cases} 
\frac{1}{2}(\textbf{a} - \hat{\textbf{a}})^2 & \text{if } |\textbf{a} - \hat{\textbf{a}}| \le \delta \\
\delta |\textbf{a} - \hat{\textbf{a}}| - \frac{1}{2}\delta^2 & \text{otherwise}
\end{cases}
\end{eqnarray}
where $\delta$ is a threshold parameter.

The training framework incorporates both conventional one timestep supervision and multi-step supervision strategies. The latter can be effectively used to both train a simulator from scratch and as a finetuning tool. During multi-step training, a scheduling approach was utilized, where the number of consecutive rollout steps progressively increased from 1 up to a predefined maximum limit during the early epochs. The autoregressive training loop is dynamically coupled with the barostat algorithm (Section \ref{sec::barostat}). At each rollout step, the simulation box dimensions $L_x$ and $L_y$ are updated to maintain the target pressure. Edge attributes and per-particle forces are iteratively recomputed based on the evolving box dimensions before being fed back into the model for the subsequent prediction step.

\section{GNN Simulator model with padded input}\label{sec::padding_model}

As an alternative to the simulator cascade resented in the main text we have also developed a padded model approach. This approach uses a single GNN simulator with an architecture identical to the one described in Section \ref{sec::gnn-simulator}, with the main difference being the input graph $G$ node features $x$. In addition to the temporal context of length $h$ consisting of current and previous step particle velocities $(v^t, v^{t+1}, \dots, v^{t-h})$, we add current timestep positions $\textbf{r}^t$, as well as the timestep index $t$. This allows the model to bridge the initialization gap by starting from a static configuration at $t=0$. 

\begin{figure}[h]
    \centering
    \includegraphics[width=6.4in]{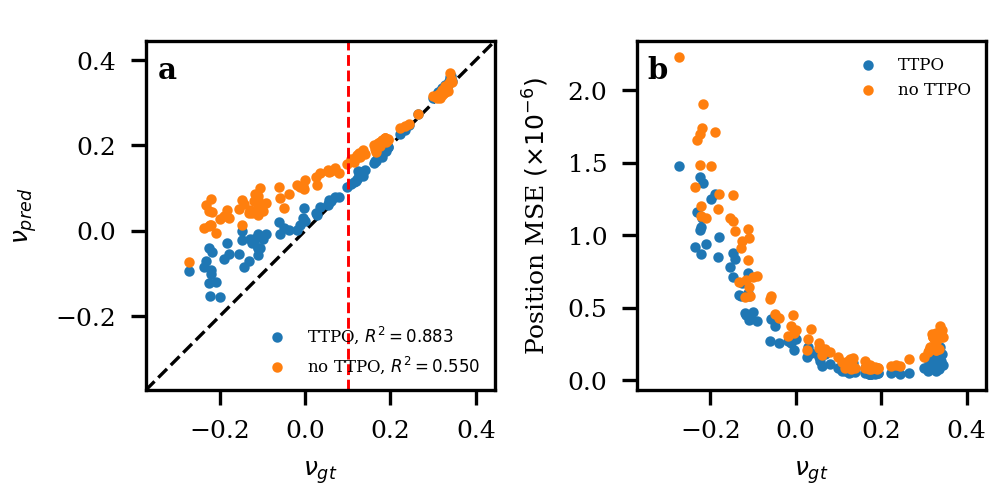}
    \caption{(a) - Parity plot showing the performance of padded GNN simulator with (blue) and without (orange) inference-time physics-based optimization. (b) - Mean-squared error between predicted and ground-truth positions at the end of a $t=50$ rollout as a function of Poisson's ratio $\nu$. The model was trained strictly on data with $\nu \geq 0.1$ shown by the dashed red vertical line.}
    \label{padding_0.1}
\end{figure}

\begin{figure}[h]
    \centering
    \includegraphics[width=6.4in]{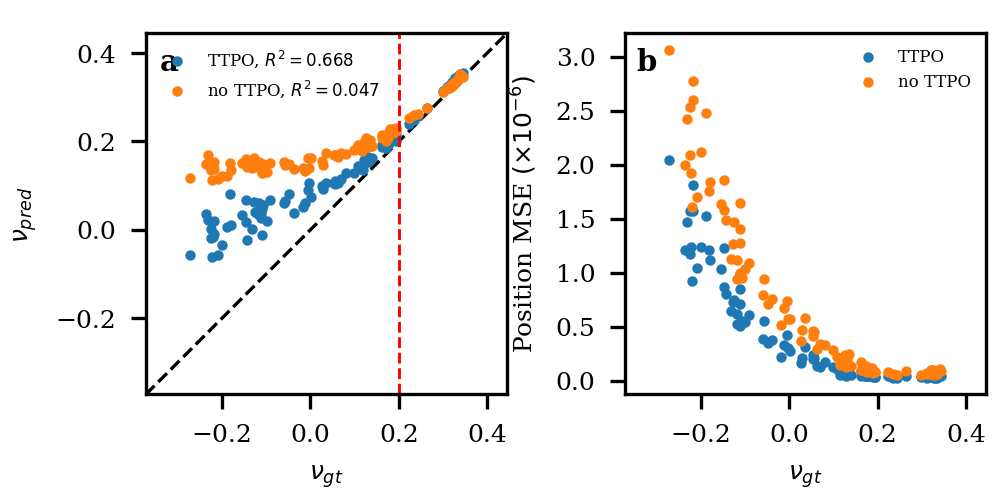}
    \caption{(a) - Parity plot showing the performance of padded GNN simulator with (blue) and without (orange) inference-time physics-based optimization. (b) - Mean-squared error between predicted and ground-truth positions at the end of a $t=50$ rollout as a function of Poisson's ratio $\nu$. The model was trained strictly on data with $\nu \geq 0.2$ shown by the dashed red vertical line.}
    \label{padding_0.2}
\end{figure}

In the beginning of a rollout when $t < h+1$, the temporal context is incomplete. To address this, the velocity part of the input tensor is padded with zeros (hence the name). As $t$ increases, the input graph node features are populated with true historical particle velocity features $v^t=r^t-r^{t-1}$ up to a maximum of $h$ automatically. By utilizing the multi-step training and test-time physics-based optimization, we have managed to improve the model's performance, as can be seen in Figures \ref{padding_0.2} and \ref{padding_0.1}.

\subsection{Padded simulator training}

The training procedure for the padded GNN simulator is identical to the one used for the history GNN simulator (Section \ref{sec::gnn_simulator::training}). To give the model the baseline understanding of how a physical systems at low timestep $t$ looks like, multi-step training included both inputs constructed from the model's previous output configurations as well as the ones constructed from the ground truth configurations. 

\clearpage
\section{Simulator cascade}\label{sec::combined_models}

Within the simulator cascade approach, an ensemble of $h$ individual GNN simulators is used to predict the compression trajectory starting from a static configuration. The cascade consists of a sequence of models with increasing input history requirements, starting at $h=0$. For the base model ($h=0$), the node features of the input graph $G^{h=0}$ contain only the network's bead positions $\textbf{r}$. For the subsequent models ($h \ge 1$), bead velocity is added to the input graph node features, calculated as $v^t = r^t - r^{t-1}$. Each subsequent model utilizes a larger temporal context window $(v^t, v^{t-1}, \dots, v^{t-h})$. During the initial phases of a rollout, this allows the cascade to build the required temporal context window, circumventing the initialization gap that occurs when starting from a single configuration at $t=0$.

\subsection{Simulator cascade training}\label{sec::combined_models::training}

The training process for the simulator cascade follows a two-stage procedure: sequential progressive training followed by end-to-end multi-step refinement.

In the first stage, each model in the cascade is trained sequentially. When training a target model with index $k$, all preceding models ($0$ to $k-1$) are instantiated and their weights are frozen. To generate the training inputs for the $k$-th model, the frozen $k-1$ models perform a short rollout starting from a ground-truth static configuration. At each step of this frozen rollout, the periodic box dimensions are dynamically updated using the pseudo-barostat algorithm. The $k$-th model then receives this generated trajectory as its input history and is trained to make a single-step prediction to match the ground-truth target. This approach exposes the target model to the accumulated prediction errors of the previous models, teaching it to inherently correct for deviations early in the trajectory.

The second stage of the training procedure is an end-to-end multi-step refinement run. During this phase, all models in the cascade are unfrozen and their parameters are chained to a single Adam optimizer with a reduced learning rate ($5 \times 10^{-5}$). The training loop simulates continuous rollouts for a predefined number of steps. At each integration step, the algorithm dynamically selects the appropriate model from the cascade based on the currently available history length. The Huber loss is calculated at each step between the active model's prediction and the ground truth. Crucially, the gradients flow across the entire unrolled cascade of different models. Gradient clipping with a maximum norm of 1.0 was used. This end-to-end refinement is essential, as it allows the sequential models to adapt, minimizing the compounding error.

\section{Minimal differentiable MD simulator}\label{sec::torch-sim}

We implemented a custom, fully differentiable molecular dynamics (MD) engine natively in PyTorch. The simulator is designed to perform uniaxial compression along the $x$-axis while allowing the $y$-axis to dynamically fluctuate to maintain a target pressure.

Compression in the $x$-direction is driven deterministically at a constant engineering strain rate $\dot{\epsilon}$. At each time step $t$, the box dimension $L_x$ is updated according to $L_x(t) = L_x(0)(1 - \dot{\epsilon} t)$, and the $x$-coordinates of all particles are scaled accordingly.

To maintain the target pressure $P_{\text{target}}$ in the transverse $y$-direction, the simulator couples the box dimension $L_y$ to a dynamic barostat. The instantaneous internal stress in the $y$-direction, $P_{yy}$, is computed using the virial theorem:
\begin{eqnarray}
P_{yy} = \frac{1}{A} \left( \sum_{i=1}^{N} m v_{i,y}^2 + \sum_{i < j} F_{ij,y} r_{ij,y} \right),
\end{eqnarray}
where $A = L_x L_y$ is the current box area, $m$ is the particle mass, $v_{i,y}$ is the $y$-component of the velocity, and $F_{ij,y}$ and $r_{ij,y}$ are the $y$-components of the pairwise force and distance vectors, respectively. The barostat introduces a box velocity $v_{b,y}$ which is driven by the pressure differential $\Delta P = P_{yy} - P_{\text{target}}$. The inertia of the barostat is governed by a piston mass $W_y$.

The equations of motion are integrated using a half-step Verlet algorithm with a timestep $\Delta t = 0.01$. A Langevin thermostat is coupled to the particle velocities, introducing a Gaussian random force $\textbf{F}_{\text{rand}} \sim \mathcal{N}(0, \sigma^2)$ and a frictional drag proportional to the particle velocity and a damping parameter $\gamma$.

A single integration step works as follows:
\begin{enumerate}
    \item The affine deformation along the $x$ axis is applied. The internal forces, virial stress, and $P_{yy}$ are evaluated at the scaled coordinates.
    \item The box velocity $v_{b,y}$ is advanced by $\Delta t / 2$ based on the pressure differential. Particle velocities are subsequently advanced by $\Delta t / 2$, incorporating the deterministic bond forces, the Langevin thermostat (friction and noise), and a metric tensor scaling factor $\exp(-v_{b,y} \Delta t / 2)$ that couples the particle velocities to the expanding/contracting box.
    \item The transverse box dimension $L_y$ and the $y$-coordinates of all particles are propagated forward by the full timestep $\Delta t$ using the half-step velocities.
    \item Forces and pressure $P_{yy}$ are re-evaluated at the newly updated positions.
    \item The box and particle velocities are advanced by the final $\Delta t / 2$ using the updated forces and pressure. 
\end{enumerate}

Following the integration step, the center-of-mass momentum is zeroed to prevent drift.

\clearpage
\section{One-step vs. multi-step supervision}

\begin{figure}[htbp]
    \centering
    \includegraphics[width=5.0in]{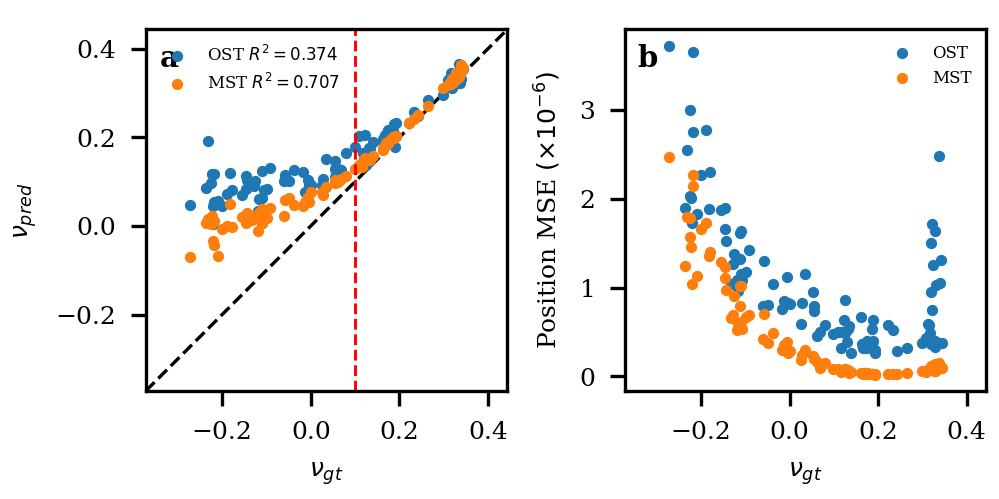}
    \caption{Comparison between two GNN simulator models trained using a one-step training (OST) strategy (blue) and multi-step training (MST) strategy (orange) on data with Poisson's ratio $\nu > 0.1$ (red dashed line). Rollouts were generated using a transient bootstrap trajectory produced by the differentiable MD engine.}
    \label{fig::MST_vs_OST_a}
\end{figure}

\begin{figure}[htbp]
    \centering
    \includegraphics[width=5.0in]{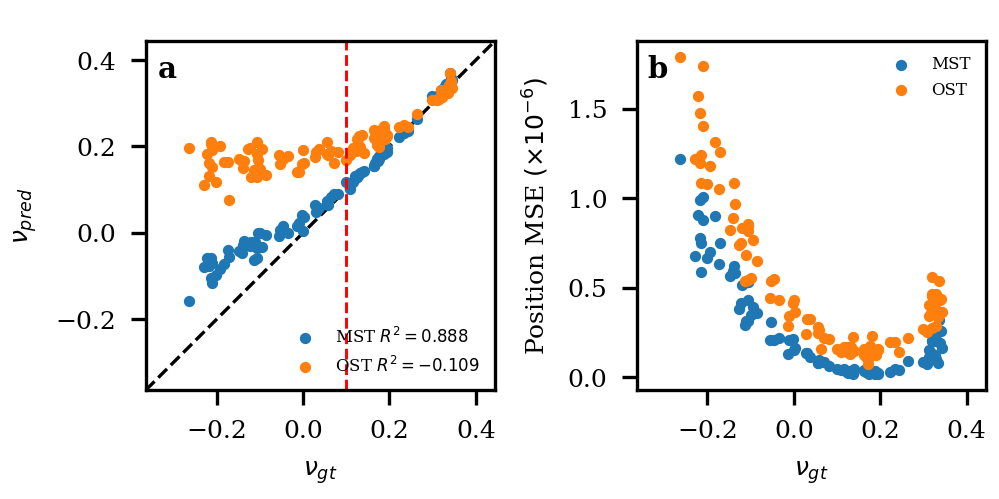}
    \caption{Comparison between two simulator cascades trained using a one-step training (OST) strategy (blue) and multi-step training (MST) strategy (orange) on data with Poisson's ratio $\nu > 0.1$ (red dashed line).}
    \label{fig::MST_vs_OST_b}
\end{figure}

\clearpage
\section{Inference-Time Physics-based Optimization}

\subsection{Physics constraints}

The net force on each particle $\mathbf{F}_i$ is derived from the harmonic bond stiffness $k = \frac{1}{l_0}$, where the total force on each particle is:
\begin{eqnarray}
    \mathbf{F}_i = \sum -k (l - l_0) \frac{\mathbf{r_{ij}}}{l},
\end{eqnarray}
where $l$ and $l_0$ are the current and rest bond lengths, while $\mathbf{r_{ij}}$ is the bond vector. The force loss component penalizes any non-zero forces along the transverse axis $y$ across the $N$ particles and is defined as:
\begin{eqnarray}
    \mathcal{L}_{\text{force}} = \frac{1}{N} \sum (F_{i, y})^2
\end{eqnarray}
The potential energy constraint $\mathcal{L}_U$ is defined as a sum of individual harmonic bond energies $E(\textbf{r}_{ij})$ (Equation \ref{eq::harmonic_energy}):
\begin{eqnarray}
    \mathcal{L}_U = \sum E(\textbf{r}_{ij})
\end{eqnarray}
The pressure loss term is defined as the squared virial stress:
\begin{eqnarray}
    \mathcal{L}_{\text{pressure}} = (\sigma_{yy})^2
\end{eqnarray}
where $\sigma_{yy}$ is computed using the pairwise harmonic forces $\mathbf{f}_{ij, y}$ and edge vectors $\mathbf{r}_{ij, y}$ over area $A$:
\begin{eqnarray}
   \sigma_{yy} = \frac{1}{2A} \sum \mathbf{f}_{ij, y} \cdot \mathbf{r}_{ij, y} 
\end{eqnarray}

\subsection{Empirical weights tuning}

Maximizing itpo performance requires carefully tuning the individual weights $\alpha_i$ of the physics constraints:
\begin{eqnarray}
    \mathcal{L}_{\text{total}} = \mathcal{L}_{\text{anchor}} + \sum_i \alpha_i \mathcal{L}_i^{\text{phys}},
\end{eqnarray}
This presents a certain methodological challenge. Because our goal is OOD generalization to unseen dynamics ($\nu < 0.1$), we cannot utilize these potentially unavailable data to tune the ITPO weights. However, if the underlying GNN simulator has already been trained using the highly stable multi-step strategy on data extending down to $\nu = 0.1$, it generalizes to this validation set \textit{too well}. Consequently, the hyperparameter search converges on physics weights of virtually zero $\alpha_i \approx 0$, as the uncorrected GNN predictions are already sufficient. This leaves the refinement loop effectively disabled for later inference on true OOD data with $\nu < 0.1$. To bypass this and ensure the physics constraints remain active, we utilize a staggered training and tuning pipeline:
\begin{enumerate}
    \item First, a GNN simulator model or a simulator cascade is trained strictly on "in distribution" data with $\nu > 0.2$ using the multi-step training strategy.
    \item Next, an automated hyperparameter optimization for itpo physics constraints weights $\alpha_i$ is performed using the intermediate finetuning data with $\nu \in [0.1, 0.2]$. Because the model has not yet seen this dynamics regime, optimizer relies on the physics constraints to minimize the loss, yielding valid, non-zero weights.
    \item Finally, the model used for the search is further trained on the intermediate data with $\nu \in [0.1, 0.2]$ using a multi-step training strategy.
\end{enumerate}

To optimize individual physics constraints weights $\alpha_i$ we employ the Optuna hyperparameter optimization framework. The objective function is defined through the coefficient of determination ($R^2$) between the predicted and ground-truth Poisson's ratios ($\nu$) over a 30-step rollout. In total, 5 different parameters are optimized:
\begin{enumerate}
    \item Refinement iterations $N_{\text{iter}}$): The number of gradient descent steps performed at each rollout timestep, sampled uniformly in the range $[5, 40]$.
    \item Learning rate $\eta$ for the Adam optimizer, sampled on a log-uniform scale between $10^{-7}$ and $10^{-4}$.
    \item The scalar coefficients for the force $\alpha_{\text{force}}$, potential energy $\alpha_{\text{energy}}$, and virial pressure $\alpha_{\text{pressure}}$ terms. These are sampled log-uniformly between $10^{-7}$ and $10^{-1}$ to account for the varying orders of magnitude of the underlying physical quantities.
\end{enumerate}

200 optimization trials were performed. For each trial, a specialized rollout on a subset of the validation data is performed.

\newpage
\section{Node trajectories}\label{sec::node_traj}
\begin{figure}[h]
    \centering
    \includegraphics[width=6.4in]{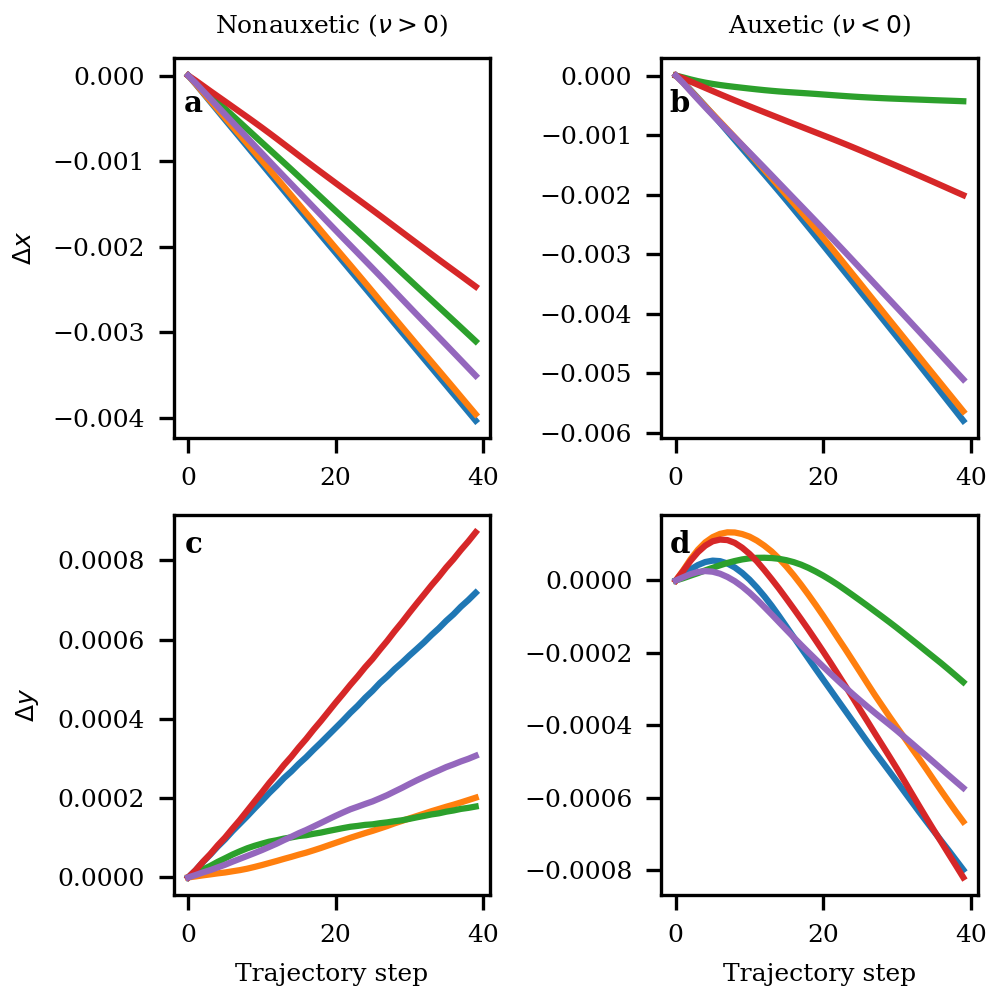}
    \caption{$x$ (a, b) and $y$ (c, d) components of selected node trajectories from nonauxetic (a, c) and auxetic (b, d) disordered elastic networks optimized via global node optimization strategy described in Section \ref{sec::opt::global_node}.}
    \label{}
\end{figure}

\newpage
\begin{figure}[ht]
    \centering
    \includegraphics[width=6.4in]{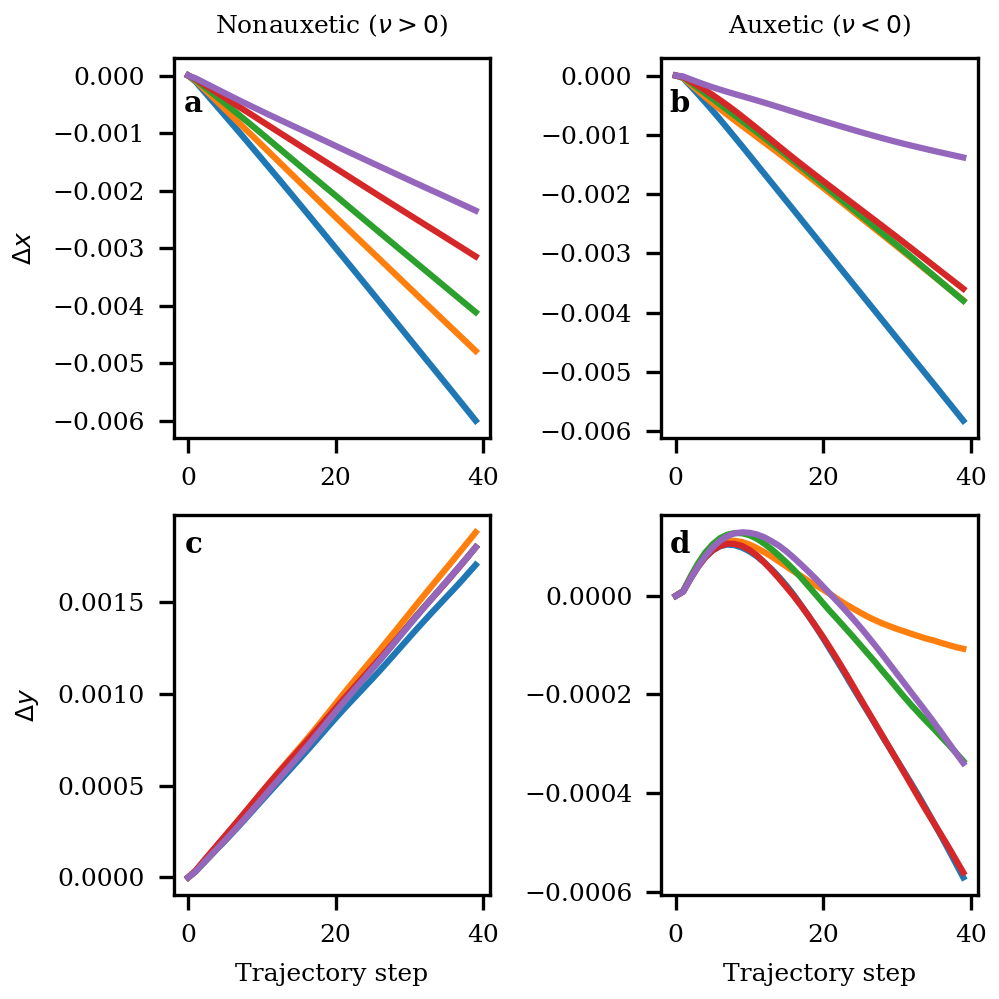}
    \caption{$x$ (a, b) and $y$ (c, d) components of selected node trajectories from nonauxetic (a, c) and auxetic (b, d) disordered elastic networks optimized via stiffness-based optimization described in Section \ref{sec::opt::stiff}.}
    \label{}
\end{figure}

\clearpage
\section{Performance on DENs optimized via direct bond stiffness tuning}

\begin{figure}[h]
    \centering
    \includegraphics[width=4.5in]{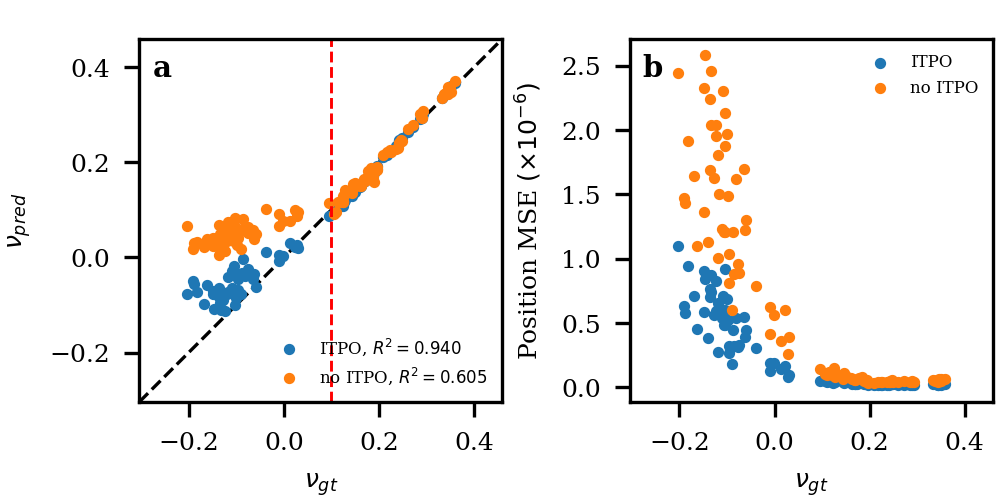}
    \caption{Performance of MD-bootstrapped GNN simulator on stiffness-tuned data with (blue) and without (orange) ITPO over 50-step rollouts. (a) Parity plot of predicted $\nu_\text{pred}$ versus ground truth $\nu_{gt}$ Poisson's ratio. (b) Mean-squared position error ($\times 10^{-6}$) as a function of $\nu_{\text{gt}}$ Dashed black and red lines represent the identity $\nu_{\text{gt}} = \nu_{\text{pred}}$ and the training data cut-off at $\nu =0.1$, respectively. The ITPO weights were searched for using a one-step trained GNN simulator on data with $\nu \geq 0.25$.}
    \label{}
\end{figure}

\begin{figure}[h]
    \centering
    \includegraphics[width=4.5in]{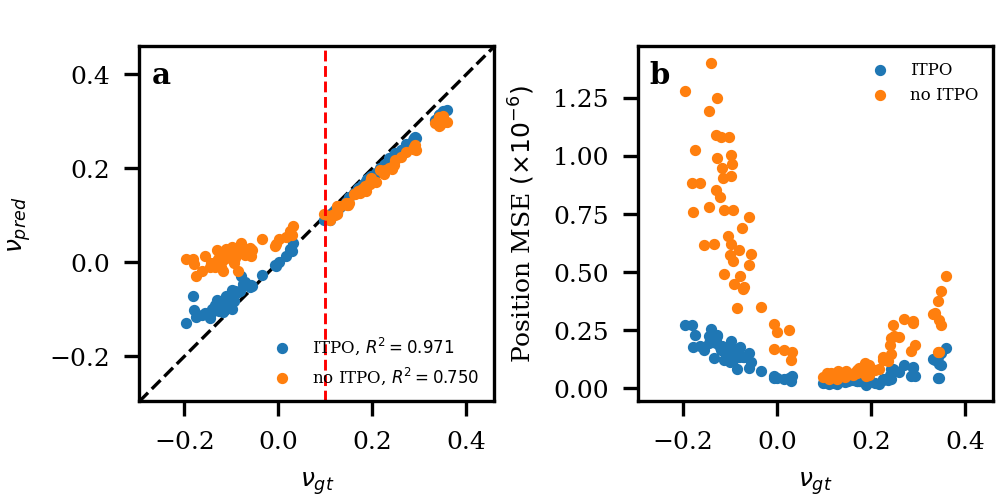}
    \caption{Performance of simulator cascade on stiffness-tuned data with (blue) and without (orange) ITPO over 50-step rollouts. (a) Parity plot of predicted $\nu_\text{pred}$ versus ground truth $\nu_{gt}$ Poisson's ratio. (b) Mean-squared position error ($\times 10^{-6}$) as a function of $\nu_{\text{gt}}$ Dashed black and red lines represent the identity $\nu_{\text{gt}} = \nu_{\text{pred}}$ and the training data cut-off at $\nu =0.1$, respectively.}
    \label{}
\end{figure}

\clearpage
\section{Further generalization}\label{sec::p>0.2}

\begin{figure}[h]
    \centering
    \includegraphics[width=6.4in]{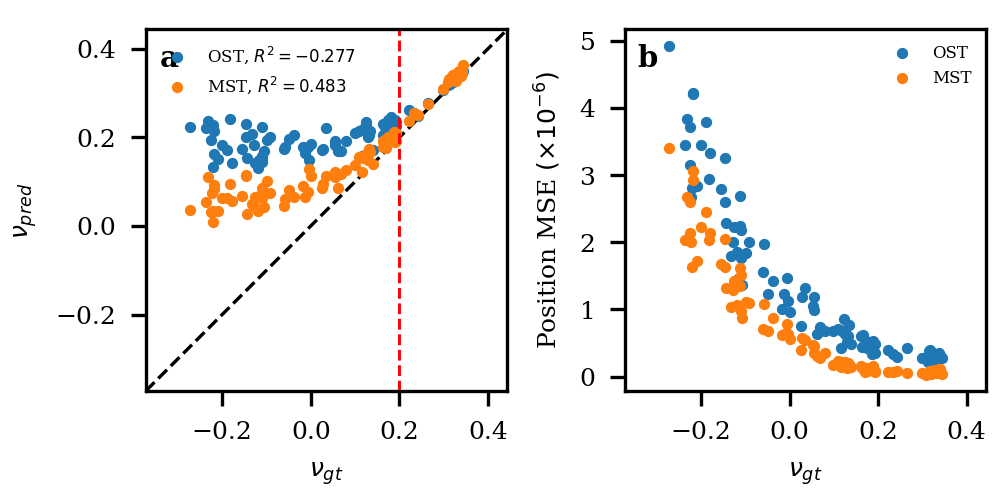}
    \caption{Parity plot (a) and mean-squared position error ($\times 10^{-6}$) as a function of $\nu_{\text{gt}}$ (b) for the MD-bootstrapped GNN simulator trained using one-step (blue) and multi-step (orange) training strategy. Dashed black and red lines represent the identity $\nu_{\text{gt}} = \nu_{\text{pred}}$ and the training data cut-off at $\nu =0.2$, respectively.}
    \label{}
\end{figure}

\begin{figure}[h]
    \centering
    \includegraphics[width=6.4in]{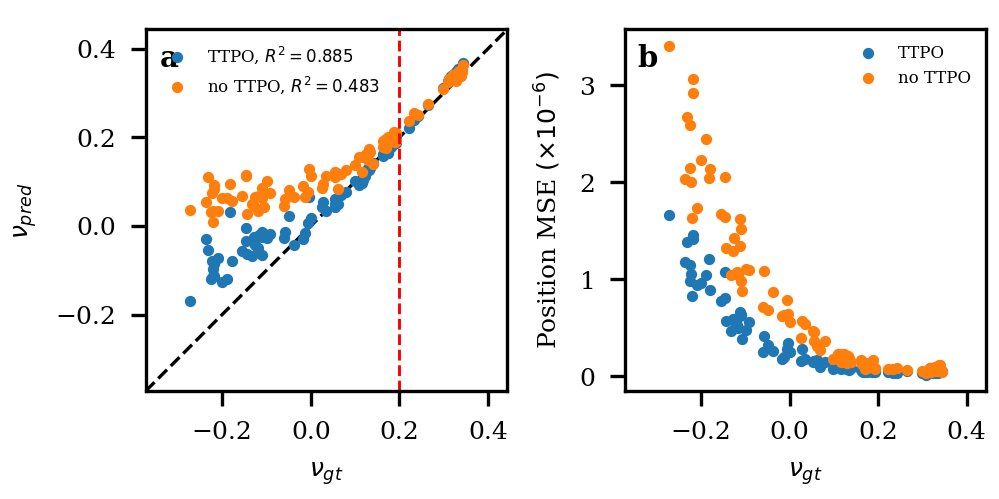}
    \caption{Parity plot (a) and mean-squared position error ($\times 10^{-6}$) as a function of $\nu_{\text{gt}}$ (b) for the MD-bootstrapped GNN simulator trained using multi-step training strategy. Blue and orange points represent results obtained with and without ITPO, respectively. Dashed black and red lines represent the identity $\nu_{\text{gt}} = \nu_{\text{pred}}$ and the training data cut-off at $\nu =0.2$, respectively.}
    \label{}
\end{figure}

\begin{figure}[h]
    \centering
    \includegraphics[width=6.4in]{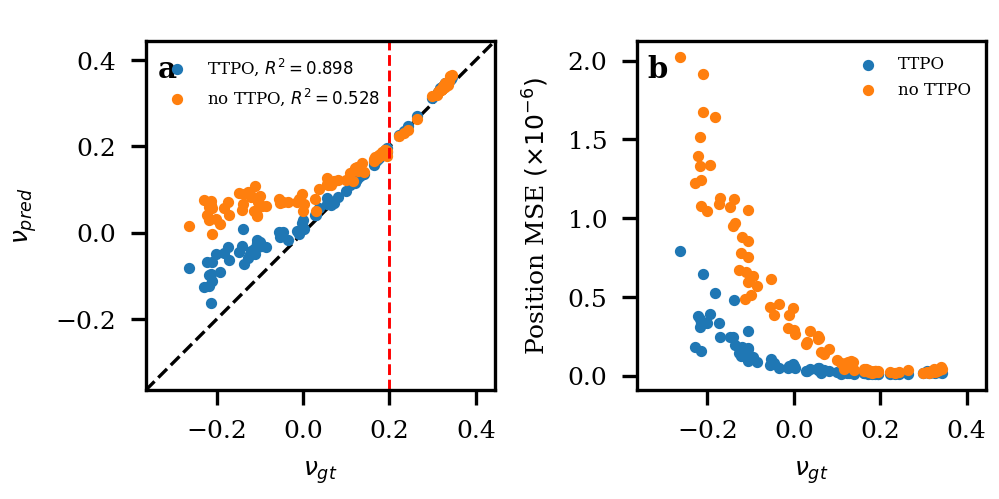}
    \caption{Parity plot (a) and mean-squared position error ($\times 10^{-6}$) as a function of $\nu_{\text{gt}}$ (b) for the simulator cascade. Blue and orange points represent results obtained with and without ITPO, respectively. Dashed black and red lines represent the identity $\nu_{\text{gt}} = \nu_{\text{pred}}$ and the training data cut-off at $\nu =0.2$, respectively.}
    \label{}
\end{figure}

\clearpage
\bibliography{papers}